\documentclass[fleqn,usenatbib]{mnras}

\usepackage[T1]{fontenc}

\DeclareRobustCommand{\VAN}[3]{#2}
\let\VANthebibliography\thebibliography
\def\thebibliography{\DeclareRobustCommand{\VAN}[3]{##3}\VANthebibliography}

\usepackage{graphicx}	
\usepackage{amsmath}	
\usepackage{amssymb}	
\usepackage{newtxtext,newtxmath}
\usepackage{multicol}   
\usepackage{makecell}   
\usepackage{caption}
\usepackage{subcaption} 
\usepackage{xurl}       

\title[$^{44}$Ti in Thermonuclear Supernova Remnants]{Upper Limits of $^{44}$Ti Decay Emission in Four Nearby Thermonuclear Supernova Remnants}

\author[J. Weng et al.]{
Jianbin Weng,$^{1}$\thanks{E-mail: jianbin.weng@smail.nju.edu.cn, pingzhou@nju.edu.cn, ygchen@nju.edu.cn}
Ping Zhou,$^{1, 2}$\footnotemark[1]
Hagai B. Perets,$^{3, 4}$
Daniel R. Wik$^{5}$
and Yang Chen$^{1, 2}$\footnotemark[1]
\\
$^{1}$School of Astronomy and Space Science, Nanjing University,
163 Xianlin Avenue,
Nanjing 210023, China\\
$^{2}$Key Laboratory of Modern Astronomy and Astrophysics, Nanjing University, Ministry of Education, China\\
$^{3}$Physics Department, Technion -- Israel Institute of Technology,
Haifa 3200003, Israel\\
$^{4}$Department of Natural Sciences, The Open University of Israel, 1 University Road, Ra'anana 4353701, Israel\\
$^{5}$Department of Physics \& Astronomy, The University of Utah, 115 South 1400 East, Salt Lake City, UT 84112, USA\\
}

\date{Accepted XXX. Received YYY; in original form ZZZ}

\pubyear{2023}

\begin{document}
\label{firstpage}
\pagerange{\pageref{firstpage}--\pageref{lastpage}}
\maketitle

\begin{abstract}
To identify progenitors and investigate evidence of He burning, we searched for decay radiation of freshly synthesized $^{44}$Ti in four young nearby thermonuclear supernova remnants: Kepler, SN 1885, G1.9+0.3 and SN 1006, by analysing the up-to-date NuSTAR archival data. No apparent flux excess from the 68 and 78 keV line emissions accompanying decay was detected above the power law continuum applied for the remnants and the absorbed stray light. By comparing the inferred upper limits of the line flux and the initial $^{44}$Ti masses with a wide variety of supernova nucleosynthesis models, we placed constraints on the supernova progenitors. We derived the first NuSTAR line flux upper limit for Kepler and ruled out most of the double-detonation scenarios with a thick He layer under low density. We estimated, for the first time, the upper limit for SN 1885, which is high because of the large distance yet still remains consistent with the He shell detonation. The new flux and mass limit of G1.9+0.3 derived from a longer total exposure is lower than the results from previous studies and evidently excludes explosive burning of He-rich matter. The relatively advanced age and the large spatial extent of SN 1006 have prevented meaningful constraints.
\end{abstract}




\section{Introduction}
\label{sec:intro}

A thermonuclear supernova (SN) is the drastic explosion of a white dwarf (WD) powered by the runaway fusion of carbon and oxygen. Thermonuclear SNe were always considered the synonym of Type Ia SNe until recent advances in SN survey revealed several subclasses with distinct photometric and spectral properties (see \citealt{2019NatAs...3..706J} for a review). It is also generally accepted that no single progenitor system or explosion mechanism could account for the whole thermonuclear SN family. Many models have been proposed to explain the distinct observational characteristics of each subtype.

The existing thermonuclear SN models can be classified based on their distinctions in specific details of the explosion. In the binary progenitor scenario, the models can be divided into the single degenerate channel, where the companion of the exploding WD is a main sequence star or an evolved star \citep{1982ApJ...253..798N, 1973ApJ...186.1007W}, and the double degenerate channel, where the companion is another WD \citep{1984ApJS...54..335I, 1984ApJ...277..355W}. Regarding the propagation of the combustion fronts, thermonuclear SN models can be characterised by detonation with supersonic shock waves \citep{1969Ap&SS...5..180A}, pure deflagration with subsonic flame fronts \citep{1984ApJ...286..644N, 2014MNRAS.438.1762F}, or their combinations, e.g., delayed detonation \citep{1991A&A...245..114K, 2013MNRAS.429.1156S} and gravitationally confined detonation \citep{2004ApJ...612L..37P,Seitenzahl2016}. Another fundamental factor for the outcome of a thermonuclear SN is the mass of the primary exploding WD, and it also changes how the fusion runaway is set off. Near the centre of a WD close to the Chandrasekhar mass limit (near-$M_{\rm Ch}$), the extreme density and temperature spontaneously lead to a runaway carbon fusion \citep{1984ApJ...286..644N}. Alternatively, a WD below the Chandrasekhar limit (sub-$M_{\rm Ch}$) can be detonated by the detonation in the accreted He shell \citep[double-detonation,][]{1990ApJ...354L..53L, 1994ApJ...423..371W, 2014ApJ...785...61S} or by the violent interactions during the merger process of the double degenerate scenario \citep[violent merger,][]{2010Natur.463...61P, 2012ApJ...747L..10P, 2013ApJ...770L...8P}. Other newly proposed methods for ignition include WD-WD head-on collisions \citep{2013ApJ...778L..37K, 2016ApJ...822...19P} and $^{235}$U fission chain reactions \citep{2021PhRvL.126m1101H}.

Some unconventional models have also been put forward for some unusual thermonuclear SNe. For example, the He shell detonation model has been proposed for Ca-rich transients, where the He shell detonation does not lead to a secondary detonation in the CO WD core \citep{2010Natur.465..322P, 2011ApJ...738...21W}. \citet{2022ApJ...924..119W} suggested the supernova remnant (SNR) G306.3-0.9 is likely the first found Galactic Ca-rich transient remnant based on its metal abundances. Another interesting model is the core-degenerate merger, which is also suggested as the progenitor of the Kepler SNR and SN PTF 11kx \citep{2013MNRAS.435..320T, 2013MNRAS.431.1541S}. In this scenario, a hot massive core of an asymptotic giant branch (AGB) star merges with its WD companion, leading to an $M_{\rm Ch}$ or super-$M_{\rm Ch}$ mass WD explosion \citep{2011MNRAS.417.1466K}.

Traditionally, the element abundances revealed by thermal X-ray spectra are used to distinguish between the characteristic nucleosynthesis of the models above. However, accurate diagnostics of various plasma parameters, such as temperature and ionisation state, are crucial for interpreting the emission lines excited by electron collisions \citep[e.g.][]{1999LNP...520..109M}. Potential large errors might arise from modelling the continuous distributions of post-shock gas properties with oversimplified discrete shock models \citep[e.g.][]{2006ApJ...645..293Z, 2009ApJ...692.1190Z, 2021ApJ...916...76A} or atomic code uncertainties \citep{2020AN....341..203M}.

In contrast, the decay radiation of the radioactive element $^{44}$Ti provides straightforward insights into its abundance and the corresponding nucleosynthesis processes. Upon de-excitation, the nucleus of $^{44}$Sc (daughter of $^{44}$Ti) emits hard X-ray line emissions at 68, 78 and 146 keV \citep{2011NDS...112.2357C}, independent of the plasma conditions. Furthermore, both the ejecta and the circumstellar medium of SNRs are optically thin to these photons.

$^{44}$Ti is mainly produced in SNe by two channels: (i) alpha-rich freeze-out, occurring when nuclear statistical equilibrium (NSE) reactions develop in a low-density regime \citep{1973ApJS...26..231W}; (ii) explosive He burning \citep{1996ApJ...464..332T}. Because few burning regions in thermonuclear SN reach the conditions of alpha-rich freeze-out, thermonuclear SNe typically produce only $\sim 10^{-6}-10^{-5}\ M_{\sun}$ of $^{44}$Ti. However, in models involving a He shell, e.g. double-detonation, the He detonation can produce a yield of $\sim 10^{-3}\ M_{\sun}$. A thick He shell detonation or a low-mass hybrid WD merger can reach a $^{44}$Ti yield up to $\sim 10^{-2}\ M_{\sun}$ \citep[for recent brief discussions, see][]{2020A&A...644A.118L, 2020A&A...638A..83W, 2023MNRAS.519L..74K}.

Because of the relatively long half-life of 59 yr, $^{44}$Ti decay radiation can power the late-time light curves and be detected in young SNRs \citep{2023MNRAS.519L..74K}. However, $^{44}$Ti decay emission from SNRs has only been robustly detected in two core-collapse SNRs, Cas A and SN 1987A, both revealing large amounts of $^{44}$Ti \citep[$\gtrsim 10^{-4}\ M_{\sun}$,][]{2014Natur.506..339G, 2015Sci...348..670B}. Such rarity is increasingly puzzling in deeper surveys, as seen in the INTEGRAL survey \citep{2006A&A...450.1037T, 2013ApJ...775...52D, 2016MNRAS.458.3411T, 2020A&A...638A..83W}. No detection in thermonuclear SNRs has been confirmed yet \citep[but see a probable \textit{Swift}/BAT detection in Tycho SNR by][which disfavoured a pure deflagration origin]{2014ApJ...797L...6T}. Nevertheless, the solar $^{44}$Ca abundance requires a moderate fraction of thermonuclear SNe to produce substantial $^{44}$Ti ejecta during the burning of He-rich material  \citep{1994ApJ...423..371W, 1996ApJ...464..332T, 2020A&A...638A..83W}, which may also account for the origin of the Galactic 511 keV emission \citep{2014arXiv1407.2254P, 2023ApJ...944...22Z}. Moreover, the inferred upper mass limits of $^{44}$Ti in SNRs can place constraints on their progenitors.

To search for possible $^{44}$Ti signatures in thermonuclear SNRs, update previous results, and provide indications on their origins, we analysed up-to-date NuSTAR archival data of four nearby targets: Kepler, SN 1885, G1.9+0.3 and SN 1006. We give brief overviews of the targets in Section~\ref{sec:targets}. The data, its reproduction and the background treatments in this study are described in Section~\ref{sec:data}. We present the results and the constraints on their progenitors in Section~\ref{sec:results} and \ref{sec:discussion}, respectively. The paper concludes with final remarks in Section~\ref{sec:conclusions}.

\section{Targets}
\label{sec:targets}

We selected targets which are positively identified as the thermonuclear type with an estimated age of $\lesssim 1000$ yr from a Galactic SNR catalogue\citep{2012AdSpR..49.1313F}. The catalogues of the Large Magellanic Cloud \citep{2016A&A...585A.162M, 2017ApJS..230....2B} and the Small Magellanic Cloud \citep{2019A&A...631A.127M} were also checked but the matching targets were not covered by NuSTAR observations. We also included SN 1885 in M31 for its peculiarity.

\subsection{Kepler}
\label{subsec:kepler}

Occurring in 1604, Kepler is the latest historical SN. Based on the recorded light curve and the dominant Fe line emission, it is one of the most certain thermonuclear SNRs \citep{2007ApJ...668L.135R, 2015ApJ...808...49K}. Oddly, the SNR is interacting with the N-rich circumstellar medium \citep{1977ApJ...218..617V, 1991ApJ...366..484B}, which suggests an AGB star wind mass-loss history and thus the single degenerate origin \citep{2010MNRAS.403.1413K, 2012A&A...537A.139C, 2019ApJ...872...45S}. However, the search for the surviving donor has failed \citep{2014ApJ...782...27K}. A possible solution is the core-degenerate scenario, where the AGB star core has merged with the WD companion near the end of the binary evolution \citep{2013MNRAS.435..320T}. The overabundance of iron-group elements (IGEs) favours the overluminous 91T-like subtype, while the fast brightness decline in Korean and European observation records is more consistent with the 91bg-like SNe \citep{2012ApJ...756....6P, 2015ApJ...808...49K, 2017hsn..book.2063V}.

No gamma rays from $^{44}$Ti decay have been confirmed in Kepler. A $2\sigma$ upper flux limit  of $1.43 \times 10^{-5}\ {\rm ph\ cm^{-2}\ s^{-1}}$ was determined for the 1157 keV $^{44}$Sc decay line from COMPTEL observations \citep{1997A&A...324..683D}. INTEGRAL/IBIS reported a $3\sigma$ upper limit of $6.3 \times 10^{-6}\ {\rm ph\ cm^{-2}\ s^{-1}}$ for the 68 and 78 keV lines \citep{2016MNRAS.458.3411T}, while INTEGRAL/SPI obtained a $2\sigma$ limit of $1.1 \times 10^{-5}\ {\rm ph\ cm^{-2}\ s^{-1}}$ for 68/78/1157 keV line combined fits \citep{2020A&A...638A..83W}. Kepler is one of the calibration sources of NuSTAR and has been monitored for over 1 Ms in 8 years, but results of a $^{44}$Ti search utilizing these data had been absent.

\subsection{SN 1885}
\label{subsec:sn1885}

SN 1885, exploding in the M31 galaxy, is one of the most recent nearby thermonuclear supernovae. Based on visual estimations in the 19th century, it exhibits many peculiar behaviours: an extremely short rise time, a fast brightness decline, a red colour throughout the evolution, etc \citep{1985ApJ...295..287D}.

Its remnant was identified as a foreground absorption spot just 16 arcsecs away from the M31 centre. The detected \ion{Fe}{i} and \ion{Fe}{ii} resonance lines revealed a Fe mass of 0.1 -- 1.0 $M_{\sun}$ , which is consistent with subluminous or normal Type Ia SNe and confirms the thermonuclear origin \citep{1989ApJ...341L..55F, 1999ApJ...514..195F, 2000ApJ...542..779H}. The HST image and spectrum analyses revealed the asymmetric \ion{Ca}{i} and \ion{Fe}{i} absorption, the outer Mg and Ca shell and the extended Fe plumes. These structures might be hints of an off-centre delayed detonation near-$M_{\rm Ch}$ explosion \citep{2007ApJ...658..396F, 2015ApJ...804..140F, 2017ApJ...848..130F}.

However, the non-detection of the remnant in the X-ray and radio band prefers a sub-$M_{\rm Ch}$ progenitor \citep{2011ApJ...730...89P, 2019ApJ...872..191S}. Moreover, the narrow light curve suggests an extremely low ejecta mass of $\sim 0.2\ M_{\sun}$ and a low explosion energy of $2.2 \times 10^{50}\ {\rm ergs}$ \citep{2011ApJ...730...89P}.

\subsection{G1.9+0.3}
\label{subsec:G1d9p0d3}

G1.9+0.3 is the youngest known SNR in the Galaxy so far \citep{2008ApJ...680L..41R}. The lack of a pulsar wind nebula or a jet-like structure, high ejecta velocities and the bilateral synchrotron X-ray rim suggest a thermonuclear origin \citep{2008ApJ...680L..41R, 2013ApJ...771L...9B}. \citet{2010ApJ...724L.161B} reported the detection of $^{44}$Sc K$\alpha$ lines ($\sim 4.1$ keV) after electron capture, translating to a 68 keV line flux of $4.7^{+3.3}_{-3.0} \times 10^{-6}\ {\rm ph\ cm^{-2}\ s^{-1}}$ with a wide line width of 4.0 keV and a $^{44}$Ti mass of $1 \times 10^{-5}\ M_{\sun}$ \citep{2013HEAD...1312706B, 2015ApJ...798...98Z}. INTEGRAL/IBIS obtained a 3$\sigma$ line flux upper limit of $9 \times 10^{-6}\ {\rm ph\ cm^{-2}\ s^{-1}}$, while NuSTAR and INTEGRAL/SPI got 2$\sigma$ upper limits of $(0.7$ -- $1.5) \times 10^{-5}\ {\rm ph\ cm^{-2}\ s^{-1}}$ and $1.0 \times 10^{-5}\ {\rm ph\ cm^{-2}\ s^{-1}}$, respectively \citep{2016MNRAS.458.3411T, 2015ApJ...798...98Z, 2020A&A...638A..83W}.

\subsection{SN 1006}
\label{subsec:sn1006}

The high Galactic latitude ($\sim 14.6\degr$) and the elemental abundances revealed by X-ray observations make SN 1006 a very likely Type Ia event \citep{2008PASJ...60S.141Y,2013ApJ...771...56U}. The failed searches for a surviving donor star disfavoured the single degenerate origin \citep{2012Natur.489..533G, 2012ApJ...759....7K}, while the discovery of a \ion{H}{i} cavity might suggest otherwise \citep{2022ApJ...933..157S}. Furthermore, the non-detection of a surviving WD indicates a merger if in the double degenerate scenario \citep{2018MNRAS.479..192K, 2022ApJ...933L..31S}. 

Previous $^{44}$Ti surveys by COMPTEL and INTEGRAL report a $2\sigma$ mass upper limit of $\sim 3 \times 10^{-2}\ M_{\sun}$ and a $3\sigma$ mass upper limit of $\sim 10^{-1}\ M_{\sun}$, respectively \citep{1999ApL&C..38..383I, 2016MNRAS.458.3411T}.

\section{Data and Methods}
\label{sec:data}

\subsection{X-ray Data and Calibration}
\label{subsec:data}

The NuSTAR observations utilised are listed in Table~\ref{tab:data}. The NuSTAR Data Analysis Software (NUSTARDAS, v2.1.1) and associated calibration database (CALDB, v20220510)\footnote{\url{https://heasarc.gsfc.nasa.gov/docs/nustar/analysis/}} files were used to reproduce data and extract spectra. We performed the spectral analysis with Xspec (v12.12.0)\footnote{\url{https://heasarc.gsfc.nasa.gov/docs/xanadu/xspec/}} and other tools in HEASoft (v6.29)\footnote{\url{https://heasarc.gsfc.nasa.gov/docs/software/lheasoft/}}.

\begin{table*}
\centering
\caption{NuSTAR Observation Log and Relevant Physical Parameters}
\label{tab:data}
\begin{tabular}{ccccccc}
\hline
Target & ObsID & Exposure $^a$ & PI & Distance & Age & Galactic Coordinates\\
     & & ks & & kpc & yr & {\it l/b}($\degr$)\\
\hline
Kepler
&   \thead{40001020002, 90201021002, 90201021004,\\90201021006, 90201021008, 90201021010,\\10501005002, 10601005002, 10701407002,\\10701407004, 10801407002, 10901407002}
&   887
&   F. Harrison
&   5.1$^b$
&   414
&   4.52/+6.82
\\
SN 1885
&   \thead{50101001002, 50201001002, 50302001002,\\50302001004, 50302001006}
&   279
&   M. Yukita
&   731$^c$
&   132
&   121.17/-21.57
\\
G1.9+0.3
&   \thead{40001015002, 40001015003, 40001015005,\\40001015007, 40702003002, 40702003004,\\40702003006, 40702003008, 40702003010}
& 811
& F. Harrison, A. Zoglauer
&   8.5$^d$
&   117$^e$
&   1.87/+0.33
\\
SN 1006
&   40110001002, 40110002002
& 381
& J. Li
&   2$^f$
&   1010
&   237.57/+14.57
\\
\hline
\multicolumn{7}{l}{$^a$ The NuSTAR/FPMA exposure after livetime corrections, data screening and removal of periods affected by solar activities}\\
\multicolumn{7}{l}{\makecell[lm{0.95\textwidth}]{$^b$ \citet{2016ApJ...817...36S}; $^c$ \citet{2015MNRAS.451..724W}; $^d$ \citet{2020MNRAS.492.2606L, 2008ApJ...680L..41R}; $^e$ \citet{2011ApJ...737L..22C}; $^f$ \citet{2013Sci...340...45N, 2002ApJ...572..888G}}}
\end{tabular}
\end{table*}

The data were reprocessed by the standard data analysis pipeline {\tt nupipeline} to generate cleaned, calibrated event list files. Strict criteria were applied to identify South Atlantic Anomaly and the ``tentacle'' passages. The periods affected by solar activities were removed by screening the light curve bins with abruptly increased count rates. The utilised total exposure time for each target is listed in Table~\ref{tab:data}.

\subsection{Spectra Extraction}
\label{subsec:region}

\begin{figure*}
\centering
\begin{subfigure}{0.24\textwidth}
    \centering
    \includegraphics[width=\textwidth]{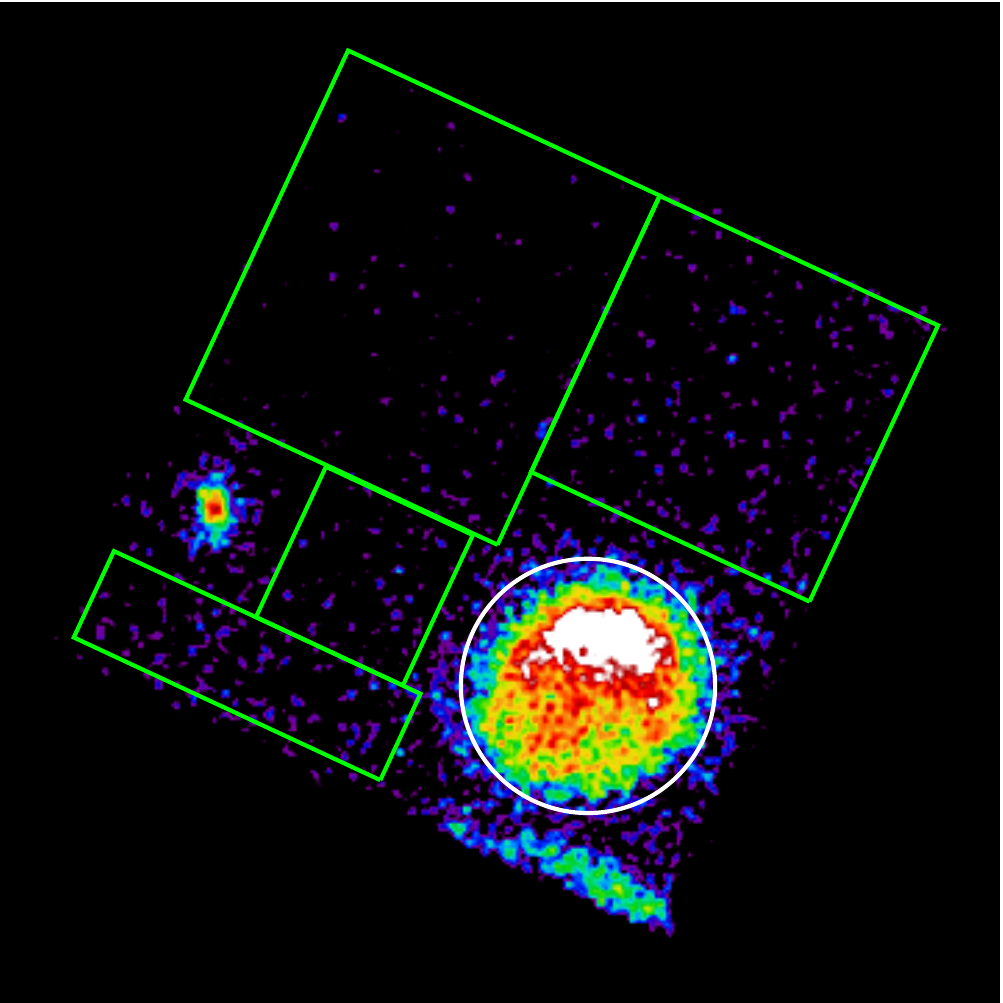}
    \caption*{Kepler (obs. 10701407004)}
\end{subfigure}
\hfill
\begin{subfigure}{0.24\textwidth}
    \centering
    \includegraphics[width=\textwidth]{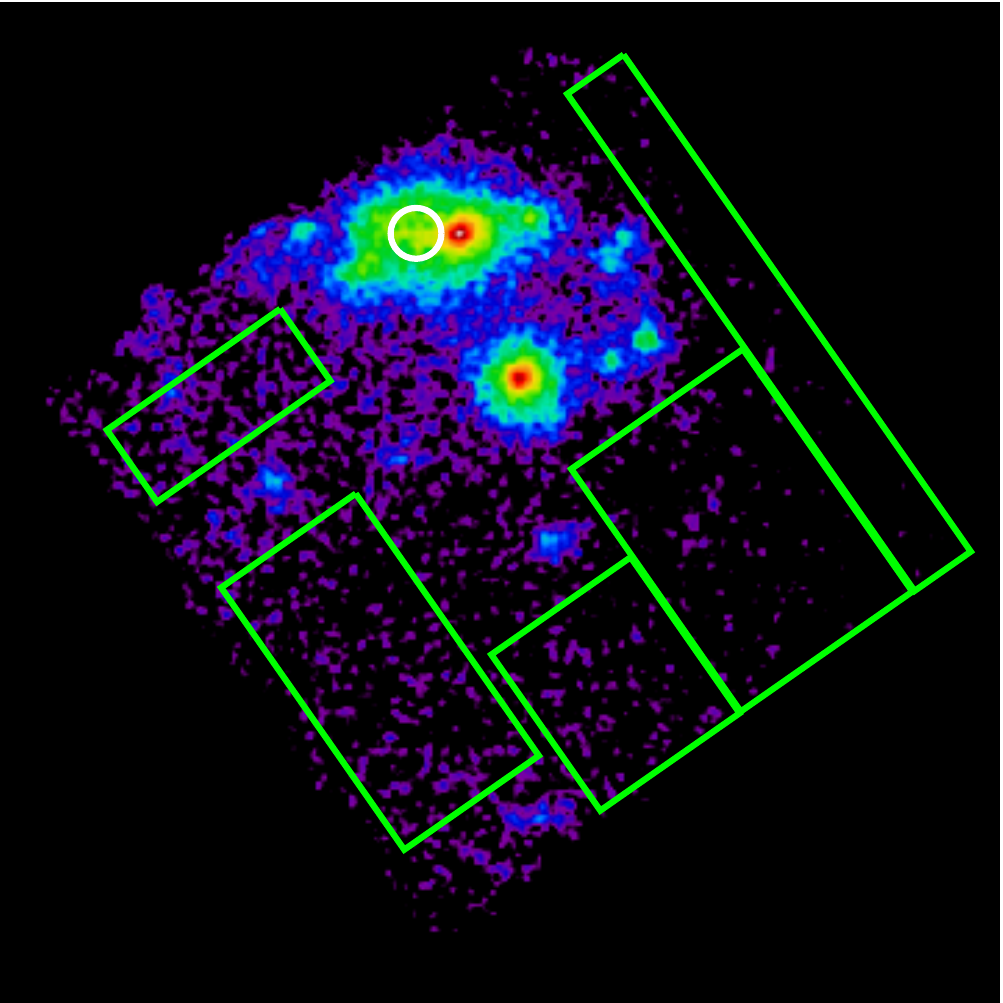}
    \caption*{SN 1885 (obs. 50201001002)}
\end{subfigure}
\hfill
\begin{subfigure}{0.24\textwidth}
    \centering
    \includegraphics[width=\textwidth]{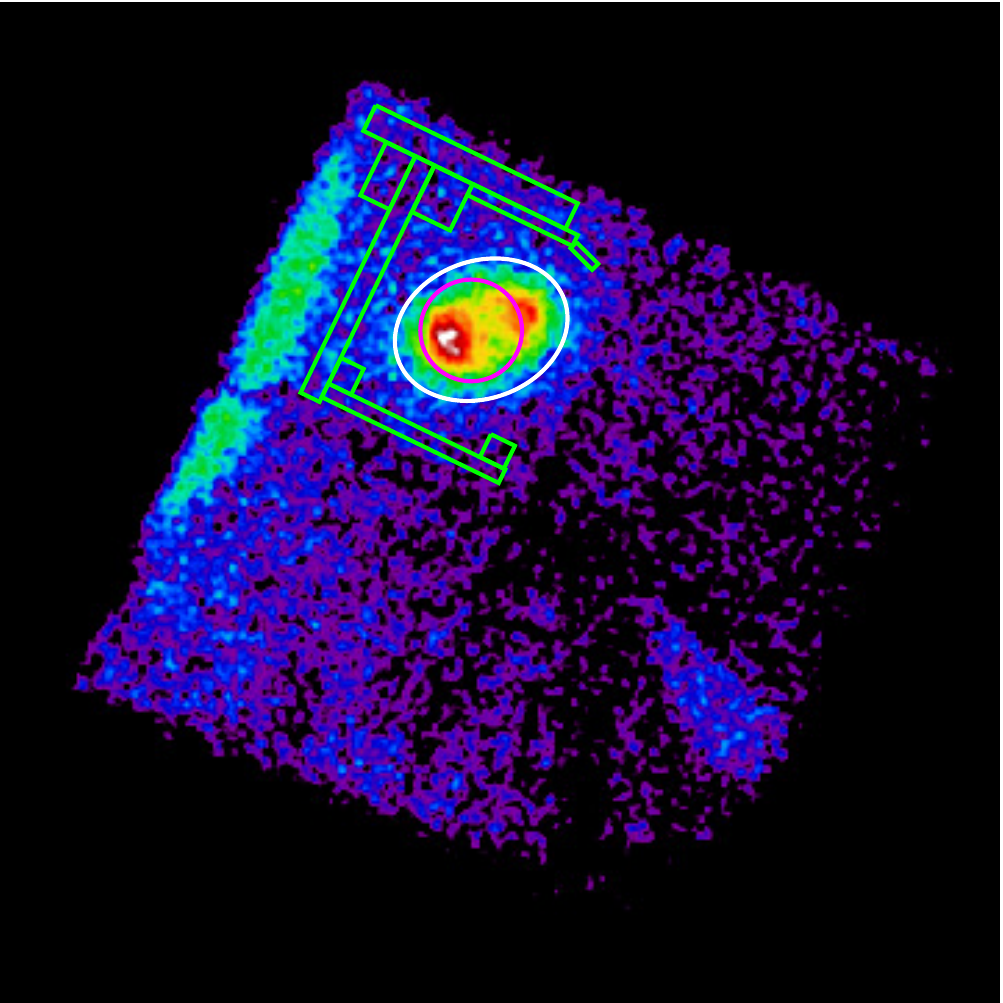}
    \caption*{G1.9+0.3 (obs. 40702003008)}
\end{subfigure}
\hfill
\begin{subfigure}{0.24\textwidth}
    \centering
    \includegraphics[width=\textwidth]{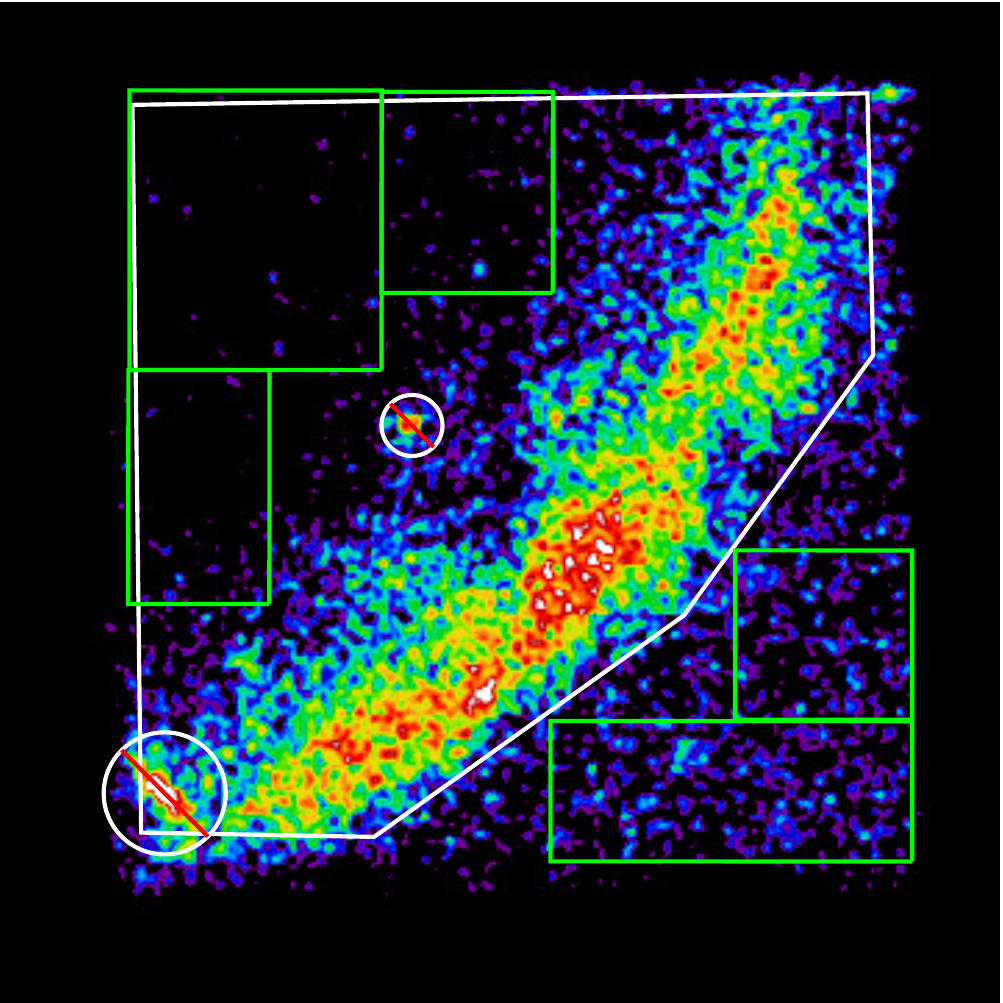}
    \caption*{SN 1006 SW (obs. 40110002002)}
\end{subfigure}
\caption{3 -- 20 keV NuSTAR/FPMB images of the targets. The source regions are marked by the white regions and the green rectangles display the defined background. The extraction region of the interior 60-70 keV excess in G1.9+0.3 is indicated by the magenta circle. The bright stripe at the southwest corner of Kepler observation is the stray light from the nearby X-ray binary GX 1+4. The stray light from GRS 1758-258 (northeast stripe) and multiple sources can also be seen in the G1.9+0.3 observations.}
\label{fig:sourcearea}
\end{figure*}

The spectra were extracted from the reprocessed event files using the standard dedicated module, {\tt nuproducts}. All targets were treated as extended sources, except SN 1885, which is considered a point source. In the case of extended sources, the auxiliary response files were weighted according to the event counts across the extended regions, thereby reflecting the spatial distribution of the response. The source regions were defined as slightly larger than the entire SNRs to account for the Point Spread Function wing, which is not included in the extended source response. We also avoided the regions illuminated by the stray light (SL), photons that reach the focal plane directly from nearby bright sources without passing through any optics for reflection. The affected regions were determined visually and from the estimated SL map in the NuSTAR Stray Light Source Catalog \citep[StrayCats,][]{2021ApJ...909...30G}. Figure~\ref{fig:sourcearea} displays the extraction regions of some observations as an example. More details can be found in Appendix~\ref{subsec:extraction}.

The source spectra were binned with the optimal binning scheme based on the response matrices \citep{2016A&A...587A.151K}. We also checked that the bins in the generated background spectra (see Section~\ref{subsec:background}) contained at least 3 photons to reduce the systematic errors that occurred when the W-statistic \citep[labelled as C-stat in Xspec,][]{1979ApJ...228..939C} is used for the spectral fitting of weak sources \footnote{\url{https://giacomov.github.io/Bias-in-profile-poisson-likelihood/}}.

\subsection{Background Subtraction}
\label{subsec:background}

For all SNRs except G1.9+0.3, we modelled the NuSTAR background with the software package {\tt nuskybgd} \citep{2014ApJ...792...48W}. The background spectra are described by the components as follows: (1) the ``Internal" background consists of a featureless continuum and a series of activation and fluorescent lines; (2) the ``Aperture" SL component is the cosmic X-ray background (CXB) which is not blocked by the aperture stops and irradiates directly the detectors; (3) the ``Solar" background comes from scattered and reflected emission from the Sun and the Earth's albedo; (4) and the ``fCXB'' component is the CXB focused by the optics. The {\tt nuskybgd} software first extracted spectra from the defined background regions with visible point sources excluded (see Figure~\ref{fig:sourcearea} for an example). Then the software performed joint fits to the spectra which include the spatial gradients of the background components. At last, {\tt nuskybgd} produced faked spectra for the source region based on the best-fit background model, which would be subtracted in later analysis.

In addition to SL, many of the observations utilised were impacted by a rare occurrence known as absorbed stray light (ASL). While photons from nearby bright sources can pass through the open gap between the aperture stops and the optical bench at small angles and become SL, photons from sources at larger angles should be blocked by the stops. Nevertheless, high-energy photons may still penetrate the blockers and reach the detector, possibly because of the absence of the Sn layer. These photons are partially absorbed and appeared as ASL, spreading over an extended area of the detector \citep[see][fig. 15]{2017JATIS...3d4003M}. We included an extra heavily attenuated power law component to address background regions affected by ASL. Detailed descriptions of the background treatments for each target and the intervening ASL are provided in Appendix~\ref{subsec:appenbkg}.

\section{Results} \label{sec:results}

The FPMA and FPMB spectra were jointly fitted with a power law model plus two velocity-shifted Gaussian lines ({\it powerlaw+vashift*(gauss+gauss)}), which describe the underlying source continuum and the $^{44}$Ti decay lines, respectively. The centroids of the Gaussian lines were fixed at 67.87 and 78.32 keV, and the fluxes were tied with the absolute $\gamma$ intensities (branching ratios) of 93.0\%/96.4\% \citep{2011NDS...112.2357C}. The velocity shifts of the lines were determined by the locations of the SNRs in Table~\ref{tab:data} and the Galactic rotation curve \citep{2018ApJ...856...52W, 2014ApJ...783..130R}. Considering the small numbers of counts in the hard end of the source spectra, all fittings were based on C-statistic. We used the Anderson-Darling statistic and the {\tt goodness} command for goodness-of-fit testing of the underlying continuum, and our corresponding models proved to be all sufficient. The scarcity of high-energy photons also meant that the dominating background Poisson fluctuations would greatly affect the line flux measurements, and hence we adopted a similar approach to \citet{2014Natur.506..339G}. We generated 1,000 faked background realizations for each spectrum and repeated the fitting to obtain a distribution of best-fit results. The mean values of the 68 keV line flux and its error were taken as the best estimates. The confidence interval was calculated by the delta statistics technique in Xspec and given at $2\sigma$ level. Further adjustments to the fitting process for each SNR are described below. The spectra and the fitting models are shown in Appendix~\ref{sec:spectra}. In all four SNRs, we did not detect well-defined $^{44}$Ti lines, and only flux upper limits were obtained.

\subsection{Kepler}
\label{subsec:keplerresult}

For observations affected by the continnum excess from ASL, a highly absorbed power law ({\it phabs*powerlaw}) was added, for which the response for SL was assigned, as in Appendix~\ref{subsec:appenbkg}. The photon indices were tied by assuming a consistent ASL spectral shape and same sources intervening the extraction regions due to comparable pointings and position angles, while those of obs. 90201021002 and 90201021004 were tied separately because the ASL was dominated by a nearby outburst event (see Appendix~\ref{sec:straylight}). The potential errors arising from spectral variability of the ASL sources are addressed in Section~\ref{subsec:error}. A cross-normalization constant was added to account for the instrument-dependent flux discrepancy. The spectral fits were performed over the 15 -- 79 keV regime. Apart from the local standard of rest (LSR) velocity of 25 km/s at a distance of 5.1 kpc \citep{2016ApJ...817...36S}, we also used the average ejecta Doppler shift of 900 km/s measured by XMM-Newton/RGS as the bulk velocity of $^{44}$Ti \citep{2021ApJ...915...42K} and got a $\sim 5\%$ smaller flux limit.

\begin{figure}
\centering
\includegraphics[width=\columnwidth]{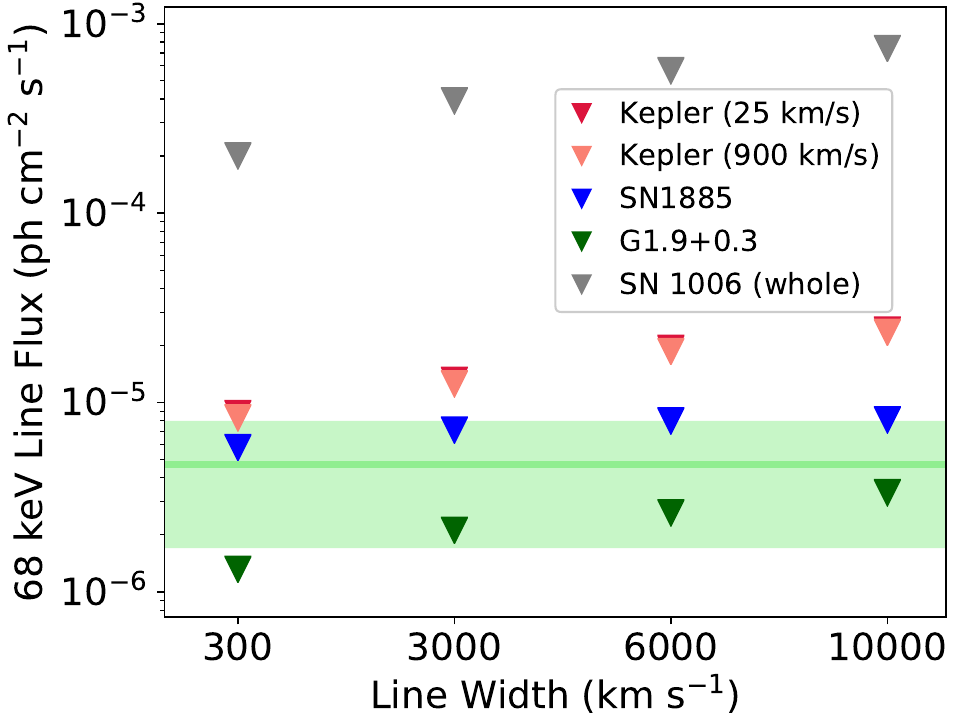}
\caption{The $^{44}$Ti 68 keV decay line flux $2\sigma$ upper limits of Kepler with bulk velocities of 25 km/s and 900 km/s, SN 1885, SN 1886 and G1.9+0.3. The light green horizontal line depicts the $^{44}$Sc 4.1 keV line flux translated from the Chandra observations of G1.9+0.3, and its 90\% confidence region is shown by the shade \citep{2013HEAD...1312706B, 2015ApJ...798...98Z}.}
\label{fig:lineflux}
\end{figure}

The line flux upper limit with a Doppler shift of 900 km/s and a width of $\sigma = 3000$ km/s is $1.3 \times 10^{-5}\ {\rm ph\ cm^{-2}\ s^{-1}}$ (see Figure~\ref{fig:lineflux}). Despite the long exposure, we obtained a consistent but higher line flux limit than that from previous INTEGRAL observations, which is likely due to the ASL.

\subsection{G1.9+0.3}
\label{subsec:G1d9result}

\begin{figure}
\centering
\includegraphics[width=\columnwidth]{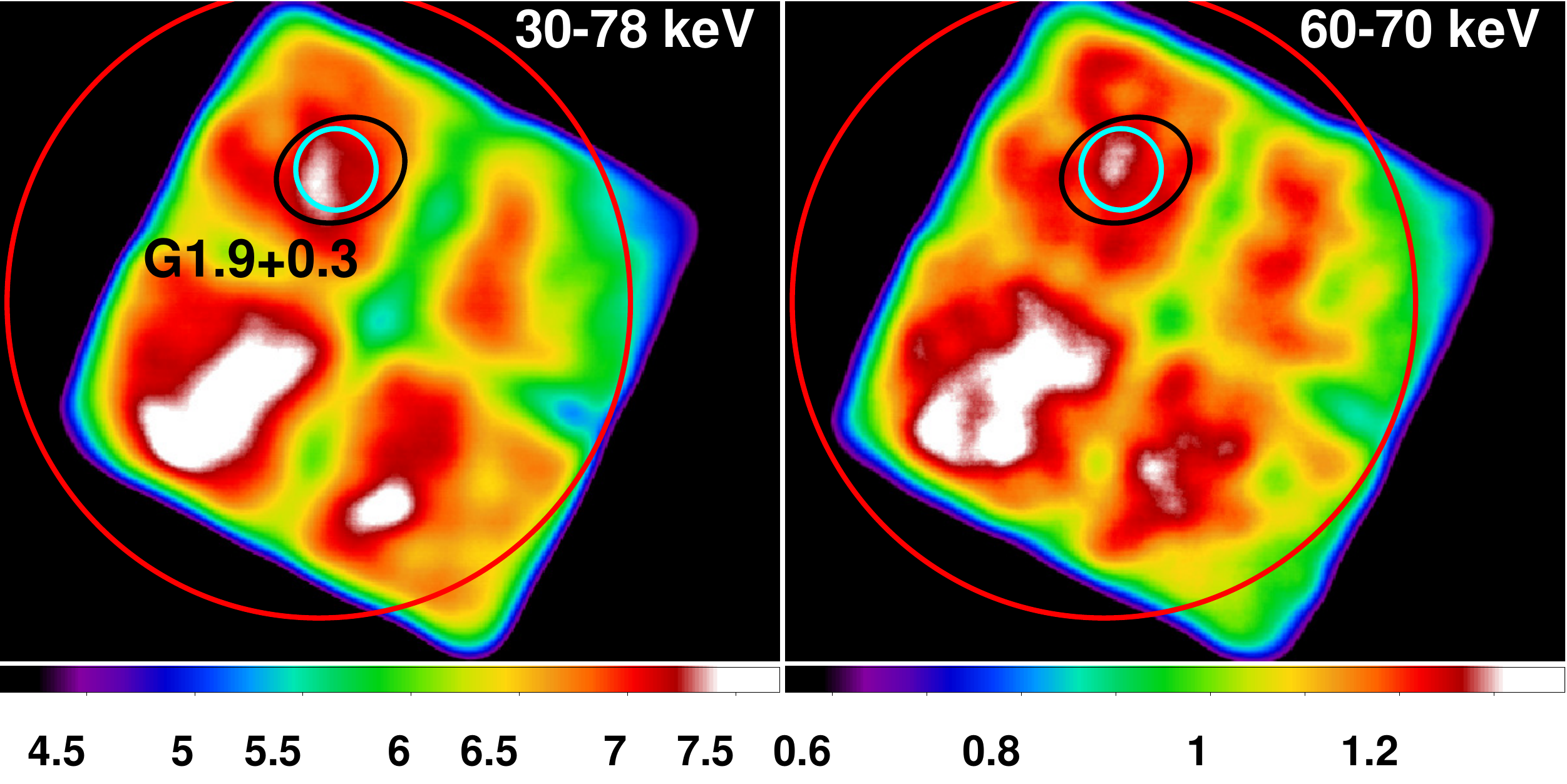}
\includegraphics[height=\columnwidth, angle=270]{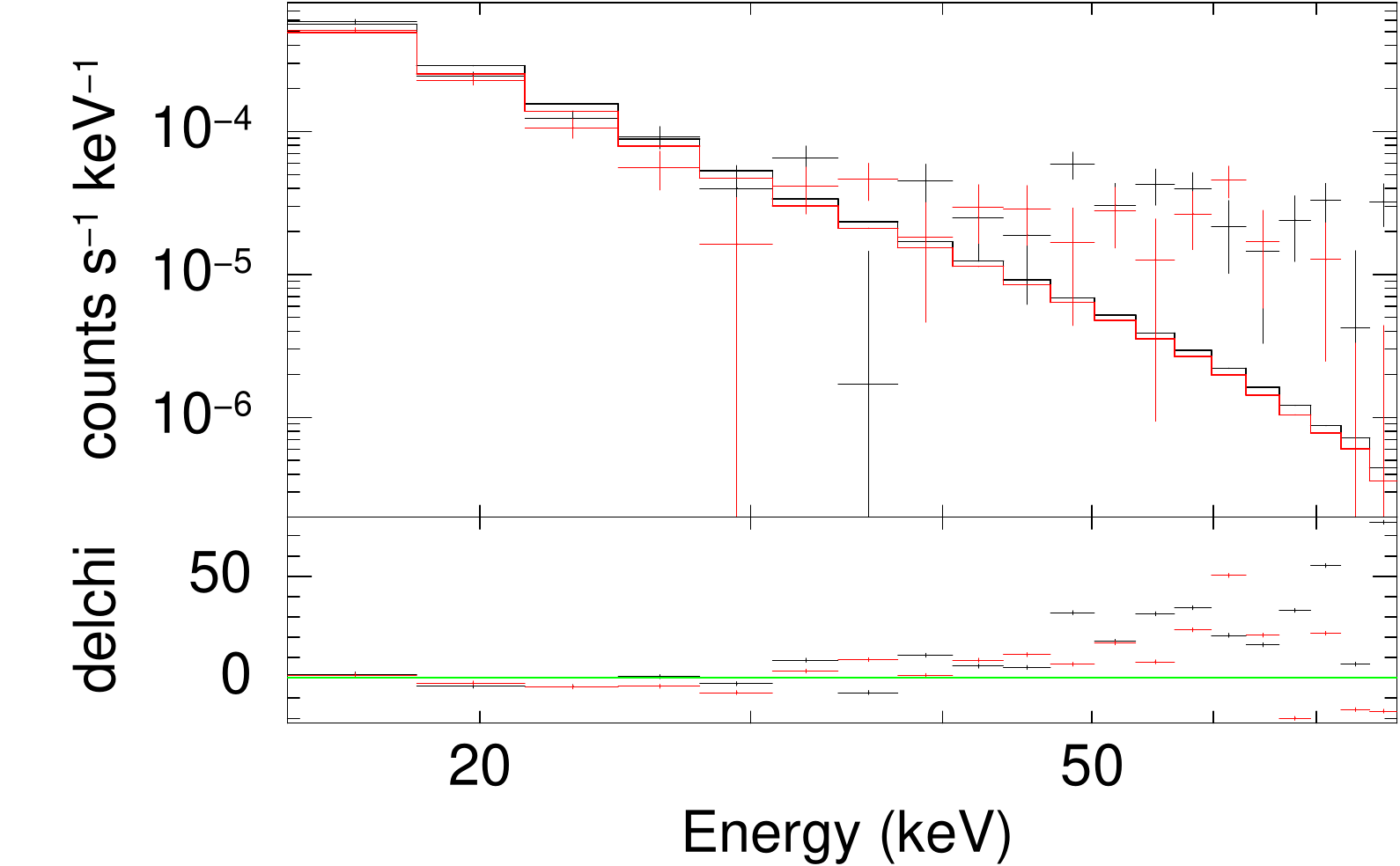}
\caption{Top: interior excess in the heavily-smoothed 30 -- 78 keV (left) and 60 -- 70 keV (right) mosaic count image of G1.9+0.3 enclosed by the cyan circle. The black ellipse depicts the source spectra extraction region in Figure~\ref{fig:sourcearea}. The arc shape of the presumed absorbed stray light (red circle) is apparent at the west half of the FoV. Bottom: G1.9+0.3 co-added FPMA (black) and FPMB (red) spectra and fitting residuals of a single power law model. The spectra show obvious residuals above 30 keV. The spectra binning displayed is for plotting purposes only.}
\label{fig:g1d9_6070}
\end{figure}

We found faint extended emissions above 30 keV in both the spectra and the combined images with a pattern similar to the predicted ASL \citep[see Figure~\ref{fig:g1d9_6070} and][]{2017JATIS...3d4003M}. It can be also seen that this excess is stronger inside the SNR and thus cannot be subtracted using background spectra from nearby regions (see high-energy spectrum residuals in Figure~\ref{fig:g1d9_6070}). Therefore, postulating an ASL origin, we adopted the same fitting procedures as in Section~\ref{subsec:keplerresult}. To bring out the faint emission, we combined the FPMA and FPMB spectra obtained from the observations conducted in July 2013 and August to October 2021, respectively (resulting in a total of four spectra after combination). We allowed the photon indices to vary, taking into account potential ASL variability. The SL response was used for the ASL component and it produced a more physical photon index ($\alpha = 4.63^{+1.41}_{-1.52}$ and $2.93^{+1.01}_{-0.65}$ for the years 2013 and 2021, respectively, with 1$\sigma$ errors) compared with the standard response ($\alpha < 0$), although the goodness-of-fit tests returned equally good results. It is noted that the photon indices presented here have limited constraints and exhibit large errors due to the scarcity of counts.

The confirmation of the ASL origin requires the identification of the SL source and its predicted ASL pattern from ray-trace simulations. The ASL shape depicted by the red circle in Figure~\ref{fig:g1d9_6070} was determined simply by eye. The focus of our analysis is to claim the necessity of a new underlying continuum, rather than the accurate fits of the ASL. 

The line flux upper limit with a width of $\sigma = 6000$ km/s is $2.6 \times 10^{-6}\ {\rm ph\ cm^{-2}\ s^{-1}}$, which is lower than previous works \citep{2016MNRAS.458.3411T, 2020A&A...638A..83W} due to the long exposure.

In the heavily smoothed combined image of 60 -- 70 keV, a seeming excess can be observed inside the SNR (Figure~\ref{fig:g1d9_6070}). We extracted spectra from an enclosing circle but still found no apparent $^{44}$Ti lines and only derived flux upper limits. Although the $^{44}$Sc line has also been found to be more prominent in the interior of the SNR \citep{2010ApJ...724L.161B, 2013HEAD...1312706B}, some excess can be seen at a similar location in the 30 -- 78 keV image. Hence this residual likely originates from the ASL rather than $^{44}$Ti decay.

\subsection{SN 1885 \& SN 1006}
\label{subsec:sn1885sn1006result}

Regarding SN 1885, the velocity shift of the line centroid was set as the radial velocity of M31 \citep[$v = -300$ km/s,][]{2012AJ....144....4M}. The spectra were fitted over 15-79 keV. The line flux upper limit with a width of $\sigma = 6000$ km/s is $8.0 \times 10^{-6}\ {\rm ph\ cm^{-2}\ s^{-1}}$ (see Figure~\ref{fig:lineflux}).

For SN 1006, due to the strong instrument lines around 20 -- 30 keV, the spectra were fitted over 35 -- 79 keV. The two NuSTAR observations only covered the northeast (NE) and southwest (SW) parts of the SNR, covering $\sim 1/3$ of the total area. The line flux upper limits with a width of $\sigma = 3000$ km/s for the NE and SW part are $7.8 \times 10^{-5}\ {\rm ph\ cm^{-2}\ s^{-1}}$ and $5.6 \times 10^{-5}\ {\rm ph\ cm^{-2}\ s^{-1}}$, respectively. Assuming uniform $^{44}$Ti distribution, the extrapolated upper limit for the whole SNR is $\sim 3.9 \times 10^{-4}\ {\rm ph\ cm^{-2}\ s^{-1}}$ (see Figure~\ref{fig:lineflux}).

\subsection{Systematic Errors}
\label{subsec:error}

The shape of the underlying continuum might significantly impact the determination of the line flux, particularly when the spectral shape is not well defined because of low counts, as is the case here. Generally, if the photon index of the ASL power law model is fixed or smaller, it will result in a lower upper limit for the line flux. This effect would be more pronounced for a broader line. To assess these errors, we set the ASL photon indices to those obtained during the background analysis of Kepler and G1.9+0.3 (see Appendix~\ref{subsec:appenbkg}). Regarding Kepler, the ASL photon index in each spectrum was fixed to the background value corresponding to the observation. This approach can also help address the potential temporal variability of the ASL, assuming the detector is solely illuminated by a single ASL source. We obtained a line flux upper limit of $1.1 \times 10^{-5}\ {\rm ph\ cm^{-2}\ s^{-1}}$ (a $\sim 10\%$ decrease). For G1.9+0.3, the ASL indices were fixed at 2.30 and 2.73 for the years 2013 and 2021, respectively, resulting in an upper limit of $2.2 \times 10^{-6}\ {\rm ph\ cm^{-2}\ s^{-1}}$ (a $\sim 15\%$ decrease). Apart from the possibility that multiple ASL sources illuminated different areas on the detector, it is noted that the ASL photon indices inferred from the background are also influenced by many factors (see Appendix~\ref{subsec:appenbkg}). Therefore, they may not necessarily provide a more accurate fit, but should be seen as a test case for errors.

Due to the absorption edge at 78.4 keV of Pt layers on the mirror modules, the response is poor near the 78 keV decay line and thus might affect the flux measurements. In some test fits, we found that removal of the 78 keV Gaussian line or the photons above 78 keV would usually lead to a $< 10\%$ difference in the acquired flux upper limits. Because the mirror response was excluded in the ASL continuum model for Kepler and G1.9+0.3, a $\lesssim 10\%$ decrease in the flux limits was obtained compared with the standard response.

The velocity shift would also affect the flux measurement, mostly due to the strong instrument line at 67.06 keV. As an extreme test, shifting the line centroid by 1,000 km/s would lead to a $\lesssim 5\%$ flux deviation for narrow line widths ($\sigma = 300$ or $3000$ km/s) and $\lesssim 3\%$ for broad line widths ($\sigma = 6000$ or $10000$ km/s).

The background spectra of G1.9+0.3 were simply scaled but not simulated repeatedly from nearby regions, and errors due to Poisson fluctuations could be large. Using the faked background spectra of Kepler, SN 1885 and SN 1006, we obtained relative standard deviations of 15 -- 30\% from the distributions of the line flux limits. The wide variance proved that it was necessary to consider the background Poisson during the spectral analysis. It is likely that the presented limits of G1.9+0.3 also hold similar errors.

The $2\sigma$ upper limits were calculated by the delta fit statistics. However, this method is only correct for the Gaussian data and is just asymptotically correct for the W-stat used in this analysis. We ran several Markov chain Monte Carlo (MCMC) test chains with steps of $6\times10^6$ (the ASL spectral shape was fixed to reduce the auto-correlation time in some of the chains) and found generally $\lesssim 10\%$ differences compared with the delta statistic technique.

To conclude, we estimated systematic errors of $\sim 15$ -- $20\%$ for Kepler, SN 1885 and SN 1006, while a larger error of $\gtrsim 25\%$ is expected for G1.9+0.3 because of the background noise and the uncertain ASL shape. We note that we did not discuss every factor (e.g. the coverage of source extraction regions or background regions), and the uncertainty on the distance also greatly affects the translation from line fluxes to $^{44}$Ti masses (see Equation~\ref{eq:mass}).

\section{Discussion}
\label{sec:discussion}

With the $^{44}$Ti half-life \citep[59.1 yr,][]{2011NDS...112.2357C}, the age and the distance to the SNR known, the decay line flux upper limit given in Section~\ref{sec:results} can be converted to the $^{44}$Ti production upper limit of the initial SN nucleosynthesis by
\begin{equation}
    \label{eq:mass}
    M(^{44}{\rm Ti}) = 44m_{\rm p} \times 4 \pi D^2 (\frac{\tau}{\ln2})2^{t/\tau} \times \frac{F_{68}}{I_\gamma}
\end{equation}
with $F_{68}$ the 68 keV line flux, $t$ the age, $D$ the distance, $\tau$ the half-life, $I_\gamma$ the absolute $\gamma$ intensity and $m_{\rm p}$ the proton mass. The calculations are based on the decay rate and absolute intensity measured in terrestrial laboratories on cold neutral $^{44}$Ti nuclide, with the modifications caused by temperature or electron states considered trivial and ignored \citep{2022hxga.book...62D}.

\begin{figure}
\centering
\includegraphics[width=\columnwidth]{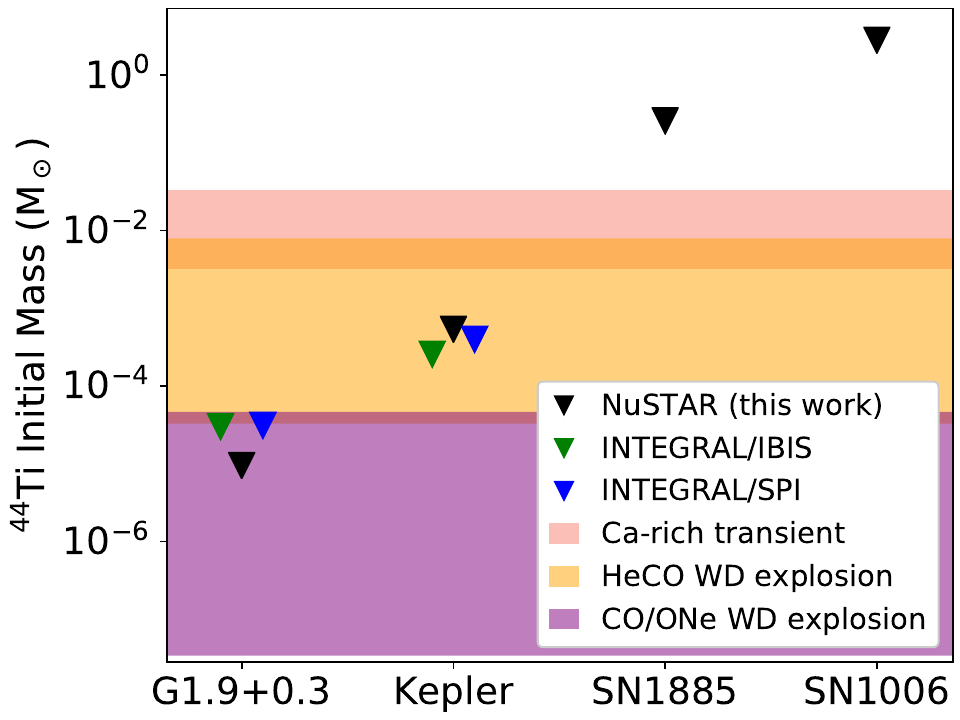}
\caption{The 2$\sigma$ upper limits of $^{44}$Ti initial mass of SNRs compared with SN theoretical yields. The INTEGRAL/IBIS (3$\sigma$) and SPI upper limits from previous works are given in green and blue triangles, respectively \citep{2016MNRAS.458.3411T,2020A&A...638A..83W}.}
\label{fig:timass}
\end{figure}

Figure~\ref{fig:timass} shows the comparisons between the derived $^{44}$Ti mass limits and the theoretical SN nucleosynthesis models listed in Table \ref{tab:model}. The ages and the distances utilised are listed in Table \ref{tab:data}. The ages were calculated by their average NuSTAR observation start dates.

\begin{table}
\centering
\caption{SN Models for Comparisons}
\label{tab:model}
\begin{tabular}{cc}
\hline
Model & Subgroup\\
\hline
Ca-rich transient
&   \thead{WD thick He shell detonation$^a$ \\ HeCO WD-CO WD merger$^b$}
\\
HeCO WD explosion
&   \thead{violent merger$^c$ \\ double detonation$^d$}
\\
CO/ONe WD explosion
&   \thead{near-$M_{\rm Ch}$ PD/DDT/GCD$^e$ \\ sub-$M_{\rm Ch}$ detonation/violent merger$^f$ \\ Iax$^g$}\\
\hline
\multicolumn{2}{l}{\makecell[lm{0.85\columnwidth}]{$^a$ \citet{2011ApJ...738...21W, 2012MNRAS.420.3003S}}}\\
\multicolumn{2}{l}{\makecell[lm{0.85\columnwidth}]{$^b$ \citet{2023ApJ...944...22Z}}}\\
\multicolumn{2}{l}{\makecell[lm{0.85\columnwidth}]{$^c$ \citet{2021MNRAS.503.4734P, 2022MNRAS.517.5260P, 2022ApJ...932L..24R}}}\\
\multicolumn{2}{l}{\makecell[lm{0.85\columnwidth}]{$^d$ The CO core might not be ignited in some models (sometimes artificially suppressed). \citet{2010ApJ...719.1067K, 2011ApJ...734...38W, 2013ApJ...774..137M, 2019ApJ...878L..38T, 2020ApJ...888...80L, Gronow2020, Gronow2021, Gronow2021b}}}\\
\multicolumn{2}{l}{\makecell[lm{0.85\columnwidth}]{$^e$ PD: pure deflagration; DDT: delayed detonation; GCD: gravitationally confined detonation. A few pure deflagration configurations might overlap the Iax models. \citet{1999ApJS..125..439I, Travaglio2004, Travaglio2005, 2010ApJ...712..624M, 2013MNRAS.429.1156S, 2014MNRAS.438.1762F, Ohlmann2014, Seitenzahl2016, 2018ApJ...861..143L, Lach2022b, 2022ApJ...925...92N}}}\\
\multicolumn{2}{l}{\makecell[lm{0.85\columnwidth}]{$^f$ \citet{2010ApJ...714L..52S, 2010Natur.463...61P, 2012ApJ...747L..10P, 2013ApJ...778L..18K, Marquardt2015, 2016MNRAS.459.4428K}}}\\
\multicolumn{2}{l}{\makecell[lm{0.85\columnwidth}]{$^g$ \citet{2015MNRAS.450.3045K, 2020ApJ...900...54L, Lach2022}}}
\end{tabular}
\end{table}

\subsection{SN Models}
\label{subsec:model}

The SN models listed in Table~\ref{tab:model} are classified according to the quantity of $^{44}$Ti production.

In the models consisting of CO or ONe WDs with no He in the initial compositions, the $\alpha$-rich freeze-out of explosive silicon burning typically produces a $^{44}$Ti yield of $\sim 10^{-6}$ -- $10^{-5}\ M_{\sun}$. We categorised these models as ``CO/ONe WD explosion".

If a hybrid WD is involved in the SN explosion, the burning of the He shell provides another channel for substantial $^{44}$Ti output. We refer to these models as ``HeCO WD explosion", consisting of WD violent merger and WD double detonation. As the fate of WDs in violent mergers can vary greatly \citep[e.g.][]{2022MNRAS.517.5260P}, they have been placed into a separate group but can also trigger double detonations in some cases. During He burning, $\alpha$ particles accumulate along the $\alpha$-chain to create new elements. Its nucleosynthesis is mainly affected by the rate of the triple-$\alpha$ reaction and subsequent $\alpha$-captures, which, in turn, is determined by factors such as the masses of the WD core/shell and the C enrichment of the shell. It can be seen in Figure~\ref{fig:timass} that the $^{44}$Ti ejecta mass of this group spans a wide range of $\sim 10^{-5}$ -- $10^{-3}\ M_{\sun}$. Under certain conditions, $^{44}$Ti would become one of the most abundant newly synthesised elements.

The $^{44}$Ti production can be further enhanced through $\alpha$-capture onto $^{40}$Ca if the SN is able to produce large amounts of Ca, resulting in a ``Ca-rich transient" (see references in Table~\ref{tab:model}). One possible scenario is when the thick He shell ($> 0.1\ M_{\sun}$) on a low-mass WD ($< 0.7\ M_{\sun}$) is detonated. The He detonation might still ignite the WD CO core in good symmetry (i.e. double detonation), producing abundance patterns different from Ca-rich transients. However, due to the He shell being the main production site, this kind of model still features comparably large $^{44}$Ti yields \citep{2012MNRAS.420.3003S}. The $^{44}$Ti overabundance can also occur in the HeCO WD-CO WD merger \citep{2023ApJ...944...22Z}, where the thinner He shell is well-mixed with the accreted CO debris before detonation. The Ca-rich transient models typically predict $\sim 10^{-3}$ -- $10^{-2}\ M_{\sun}$ of $^{44}$Ti production.

\subsection{Kepler}
\label{subsec:keplermodel}

In Figure~\ref{fig:timass}, we can see that all three upper limits exclude the He-rich models with high $^{44}$Ti outputs and Ca-rich transient models.

\begin{figure}
\centering
\includegraphics[width=\columnwidth]{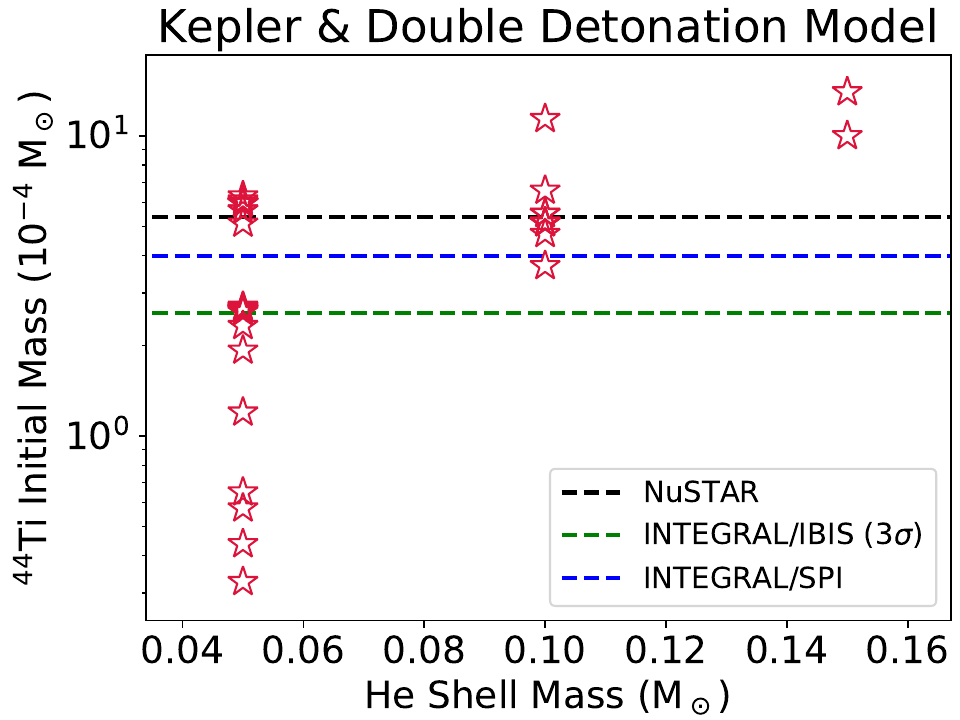}
\caption{$^{44}$Ti Yield of a group of WD double detonation models \citep[stars,][]{2020ApJ...888...80L} compared with the mass upper limits of Kepler derived from observations (dashed lines).}
\label{fig:keplerleungmodel}
\end{figure}

Figure~\ref{fig:keplerleungmodel} presents a comparison between the $^{44}$Ti productions of a set of WD double detonation models \citep{2020ApJ...888...80L} and the observational limits. The lowest limit obtained by INTEGRAL/IBIS \citep{2016MNRAS.458.3411T} reject models with thick He shells ($\gtrsim 0.1\ M_{\sun}$).

\begin{figure}
\centering
\includegraphics[width=\columnwidth]{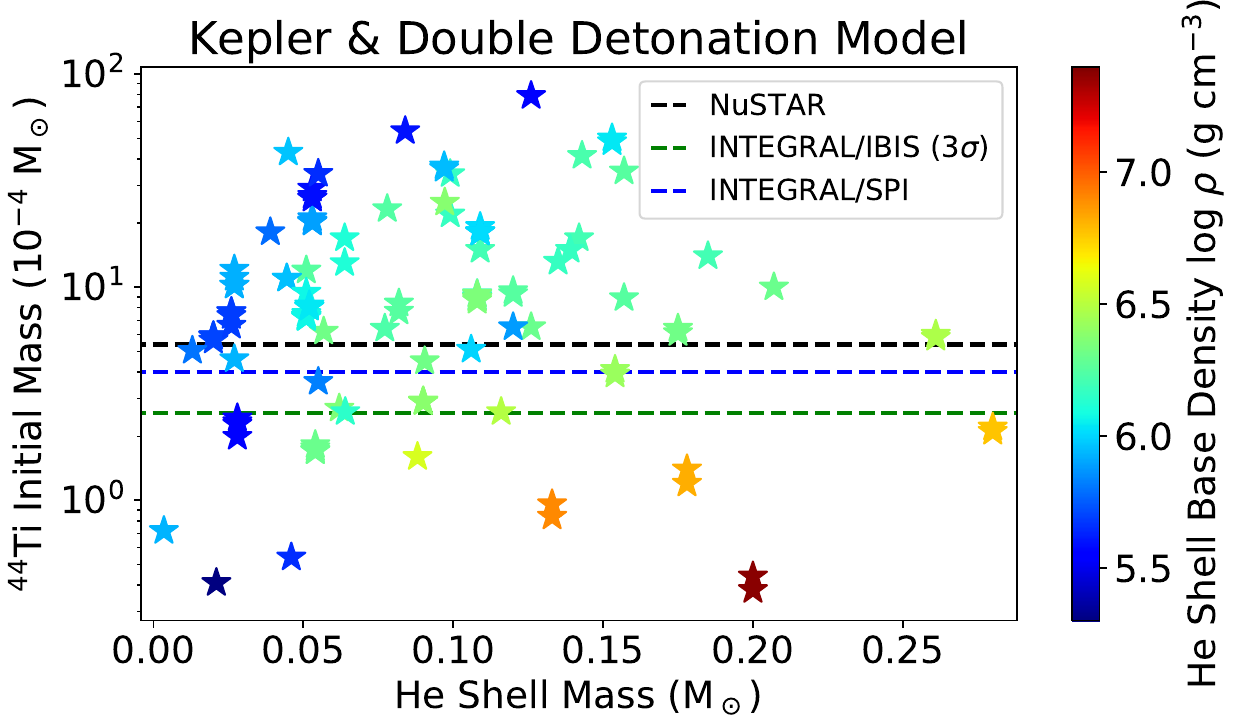}
\caption{$^{44}$Ti Yield of the WD double detonation models (stars) in Table~\ref{tab:model} with various He shell masses and He shell base densities. The mass upper limits of Kepler obtained by different telescopes are shown by dashed lines.}
\label{fig:keplermodel}
\end{figure}

However, the nucleosynthesis of He burning is also sensitive to other factors. Figure~\ref{fig:keplermodel} shows the $^{44}$Ti yields of all remaining double detonation models in Table~\ref{tab:model}, in different He shell base densities. We excluded extreme models ejecting negligible $^{56}$Ni since they cannot reproduce the IGE overabundance and the peak brightness in the historical light curve of Kepler \citep{2012ApJ...756....6P, 2015ApJ...808...49K}. It is noted that depending on specific configurations, the core-shell interface is not distinct or the runaway does not happen at the shell base in some models \citep[e.g.][]{2011ApJ...734...38W, Gronow2020}. Hence, the density depicted here is still an approximate and overgeneralized parameter. For similar densities, an increase in the $^{44}$Ti ejecta mass with the shell mass can be seen. Conversely, for a given shell mass, the $^{44}$Ti ejecta mass decreases as the density rises. Therefore, the mass upper limit can also agree with the models featuring thicker He shells but under high densities, which can be achieved through accretion onto a massive CO core unless significantly heated by nova-like events or the accretion stream \citep{2011ApJ...734...38W, 2019ApJ...878L..38T}. Another contributing factor is C enrichment in the He shell. It has been found that substantial C pollution greatly reduces $^{44}$Ti production \citep{2010ApJ...719.1067K, Gronow2020}. However, relevant calculations in thick He shells ($> 0.1\ M_{\sun}$) are still absent, and it remains unclear whether such a massive shell can survive on the WD surface \citep{2009ApJ...699.1365S}.

Among the four hybrid WD violent merger models in Table~\ref{tab:model}, only the one with thin He layers ($\sim 0.03\ M_{\sun}$ in total) is consistent with the $^{44}$Ti upper limit \citep{2022ApJ...932L..24R}. Nevertheless, the aforementioned complications may also remain relevant.

\subsection{G1.9+0.3}
\label{subsec:g1d9model}

Referring to Figure~\ref{fig:timass}, while the INTEGRAL results still reach the lower end of HeCO WD models, we infer an initial $^{44}$Ti mass limit of $9.6 \times 10^{-6}\ M_{\sun}$, which has fully entered the regime of CO WD explosions and confirms a He-poor origin.

\begin{figure}
\centering
\includegraphics[width=\columnwidth]{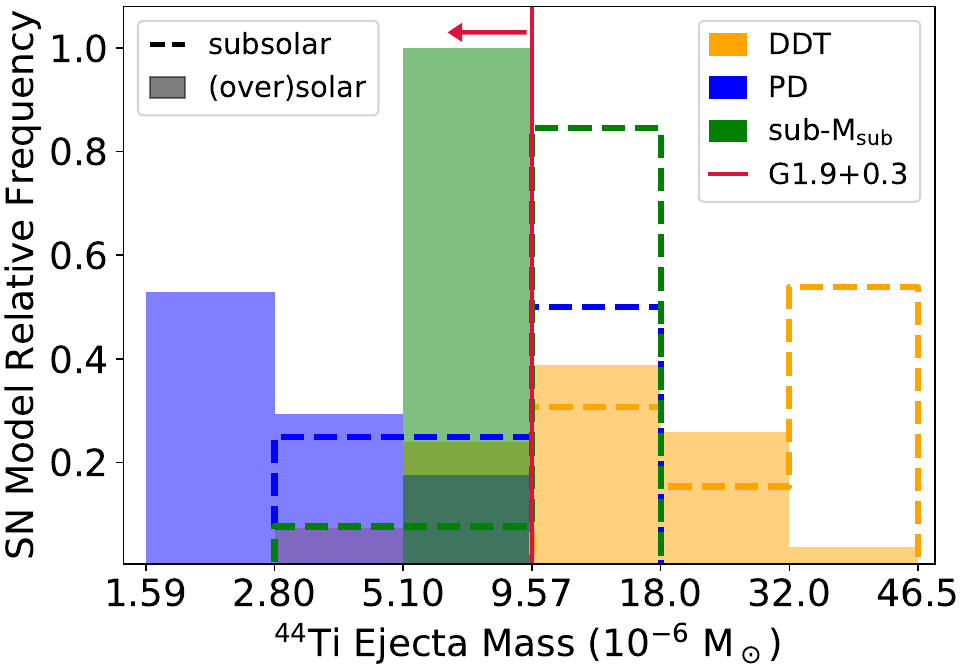}
\caption{The relative frequency of ``CO/ONe WD explosion" SN models versus their $^{44}$Ti output. Different colours correspond to different subgroups while the metallicities are depicted by shaded bars or dashed lines. The x-axis is scaled approximately logarithmically, with the lowest and highest theoretical model yield at the lower and higher ends, respectively. The vertical red line depicts the $^{44}$Ti $2\sigma$ upper limit of G1.9+0.3, translated from the line flux in Section~\ref{subsec:G1d9result}. The number of SN models depicted (oversolar/solar + subsolar): 54 + 13 delayed detonation (DDT); 17 + 4 pure deflagration (PD); 3 + 13 sub-$M_{\rm Ch}$.}
\label{fig:g1d9_snmodel}
\end{figure}

This mass limit is also lower than some CO/ONe WD models with high $^{44}$Ti yield, which provides some interesting hints to the progenitor. Figure~\ref{fig:g1d9_snmodel} shows the relative frequency of ``CO/ONe WD explosion" models listed in Table~\ref{tab:model} compared with their $^{44}$Ti yields, after excluding the models that cannot disrupt the whole WD (i.e. mass of the bound remnant $>0.05\ M_{\sun}$). We divided the models into three subgroups: pure deflagration, delayed detonation (including gravitationally confined detonation) and sub-$M_{\rm Ch}$ consisting of pure detonation and violent merger.

All pure deflagration and sub-$M_{\rm Ch}$ models with solar metallicity fall within the upper limit of G1.9+0.3 while all subsolar delayed detonation models produce more $^{44}$Ti than the limit. It is reasonable that the neutron excess brought by initial metallicity affects the production of the symmetric element $^{44}$Ti. Since a larger amount of $^{44}$Ti is expected in the delayed detonation phase \citep[e.g.][]{1999ApJS..125..439I}, it can be also seen that only a few delayed detonation but most pure deflagration models meet the mass limit. Still, an additional systematic error of 33\% or the 3$\sigma$ limit for a 10,000 km/s wide line would agree with 4 out of the 13 selected subsolar delayed detonation models for the translated mass. It is noted that the frequency in Figure~\ref{fig:g1d9_snmodel} is largely shaped by the chosen SN configurations (see references in Table~\ref{tab:model}) and should never be treated quantitatively. For example, although \citet{2013MNRAS.429.1156S} and \citet{2018ApJ...861..143L} both found that increasing metallicity would reduce $^{44}$Ti production, their different model treatments have led to slightly overlapping output ranges. Figure~\ref{fig:g1d9_snmodel} shows that the 2$\sigma$ limit of G1.9+0.3 favours an (over)solar origin and excludes the delayed detonation models with the highest $^{44}$Ti output.

The line flux upper limit obtained for a typical ejecta velocity ($> 3000$ km/s) is consistent with the 90\% confidence lower bound translated from the Sc 4.1 keV line observed by Chandra (see Figure~\ref{fig:lineflux}). However, our upper limit cannot meet the Sc 4.1 keV line best-fit flux unless the line is very wide ($> 10000$ km/s). The Chandra observation displayed a broad line width of $3.5^{+2.3}_{-2.4} \times 10^4$ km/s although being affected by the nearby Ca line \citep{2015ApJ...798...98Z}, and the width lower limit is consistent with the proper motion measurements \citep{2011ApJ...737L..22C}. However, such a velocity is too high for $\alpha$-rich freeze-out products and can be only reached by the outermost SN ejecta \citep[e.g.][]{1999ApJS..125..439I}. Although the fastest $^{44}$Ti can attain an extremely rapid velocity (up to $\sim 29000$ km/s) in certain violent mergers or gravitationally confined detonation models \citep[e.g.][]{2012ApJ...747L..10P, Lach2022b}, the majority of the $^{44}$Ti still has a speed of $\lesssim 12000$ km/s. Besides, fast-moving $^{44}$Ti ejecta seems to contradict the evidence that the Sc line is more prominent in the SNR interior \citep{2013HEAD...1312706B}. Therefore, it is probable that the true flux and width of the $^{44}$Ti decay emission in G1.9+0.3 fall within the lower confidence intervals established by \citet{2013HEAD...1312706B}, based on our upper limits. \citet{2020A&A...638A..83W} also found this minor inconsistency when comparing the Chandra signal with the upper limits they derived from INTEGRAL data. They suggested that stable isotopes like $^{45}$Sc might also contribute to this 4.1 keV fluorescent line.

\subsection{SN 1885 \& SN 1006}
\label{subsec:sn1885sn1006model}

Based on the low explosion energy and low ejecta mass \citep{2011ApJ...730...89P}, it is also likely that SN 1885 originated from a Ca-rich transient and a large amount of $^{44}$Ti might remain in the SNR after a century. However, due to the large distance to M31, for a line width of $\sigma = 6000$ km/s which is comparable to the Ca and Fe ejecta expansion velocities in optical and UV spectra \citep{2017ApJ...848..130F}, the inferred $^{44}$Ti mass limit is 0.26 $M_{\sun}$. This limit is much higher than all presented models but it also means that a Ca-rich transient origin cannot be excluded. For the NuSTAR upper limit to reach the highest $^{44}$Ti output in present Ca-rich transient models, i.e. $\sim 0.1\ M_{\sun}$ \citep{2010Natur.465..322P, 2023ApJ...944...22Z}, an additional effective exposure of at least $\sim 1.6$ Ms is required.

As for SN 1006, we derived a large $^{44}$Ti mass limit of 2.8 $M_{\sun}$ from a large line flux upper bound. It exceeds previous results obtained from the COMPTEL and INTEGRAL observations \citep{1999ApL&C..38..383I, 2016MNRAS.458.3411T}, and even a WD mass. All three results cannot provide any physical constraints to the progenitor scenario.

\section{Conclusions}
\label{sec:conclusions}

We utilised NuSTAR archival data and searched for $^{44}$Ti decay lines in the hard X-ray spectra of four nearby thermonuclear SNRs: Kepler, SN 1885, G1.9+0.3 and SN 1006. We could not find apparent decay lines at 68 or 78 keV but only obtained flux upper limits. The mass limits place some constraints on their progenitors.

1. We derived the first NuSTAR constraint on the $^{44}$Ti decay emission upper limit for Kepler, which is $1.3 \times 10^{-5}\ {\rm ph\ cm^{-2}\ s^{-1}}$. The translated $^{44}$Ti mass limit excludes the Ca-rich transient and the WD double detonation progenitor with high $^{44}$Ti output, but cannot rule out the hybrid WD explosion if the He shell is thin or under high density.

2. We obtained the first $^{44}$Ti decay line flux constraint for SN 1885. Because of the large distance, the mass upper limit of 0.26 $M_{\sun}$ is higher than the predicted yields of all current Ca-rich transient models.

3. Owing to the long exposure, the updated flux upper limit for G1.9+0.3 is lower than the limits provided by all previous observations. The mass limit of $9.6 \times 10^{-6}\ M_{\sun}$ confirms a He-poor WD origin and is lower than those predicted by the subsolar delayed detonation models. The low $^{44}$Ti ejecta mass favours a progenitor with (over)solar metallicity and a less dominant detonation phase. However, all CO WD explosion subtypes still remain possible considering systematic errors.

4. Due to the large spatial extent and old age, we were unable to obtain any physical constraints on SN 1006.

We note that the conclusions above rely on the validity of the selected SN models in Table~\ref{tab:model} and the adequacy of our crude continuum model to account for the ASL.

\section*{Acknowledgements}

This work has made use of the Heidelberg Supernova Model Archive (HESMA), \url{https://hesma.h-its.org}. This research has made use of the NUSTARDAS software, jointly developed by the ASI Space Science Data Center (SSDC, Italy) and the California Institute of Technology (Caltech, USA).  This work was made possible by the NuSTAR mission, which is led by Caltech, managed by the Jet Propulsion Laboratory, and funded by NASA.

J.W. acknowledges the help and advice on NuSTAR data reduction from Jiang-tao Li. J.W. expresses gratitude to Roman Krivonos for discussions on NuSTAR SL and ASL.

J. W., P.Z., and Y.C. thank the supports by NSFC grants 12273010, 12173018 and 12121003.
HBP acknowledges support for this project from the European Union's Horizon 2020 research and innovation program under grant agreement No 865932-ERC-SNeX.

\section*{Data Availability}

The observational data utilised in this research are publicly available through the archives of the High Energy Astrophysics Science Archive Research Center (HEASARC), accessible at \url{https://heasarc.gsfc.nasa.gov/docs/archive.html}. Interested researchers will obtain access to the intermediate and final results, as well as the corresponding analysis script employed in this study, upon reasonable request to the authors.



\bibliographystyle{mnras}
\bibliography{example} 

\begin{thebibliography}{}
\makeatletter
\relax
\def\mn@urlcharsother{\let\do\@makeother \do\$\do\&\do\#\do\^\do\_\do\%\do\~}
\def\mn@doi{\begingroup\mn@urlcharsother \@ifnextchar [ {\mn@doi@} {\mn@doi@[]}}
\def\mn@doi@[#1]#2{\def\@tempa{#1}\ifx\@tempa\@empty \href {http://dx.doi.org/#2} {doi:#2}\else \href {http://dx.doi.org/#2} {#1}\fi \endgroup}
\def\mn@eprint#1#2{\mn@eprint@#1:#2::\@nil}
\def\mn@eprint@arXiv#1{\href {http://arxiv.org/abs/#1} {{\tt arXiv:#1}}}
\def\mn@eprint@dblp#1{\href {http://dblp.uni-trier.de/rec/bibtex/#1.xml} {dblp:#1}}
\def\mn@eprint@#1:#2:#3:#4\@nil{\def\@tempa {#1}\def\@tempb {#2}\def\@tempc {#3}\ifx \@tempc \@empty \let \@tempc \@tempb \let \@tempb \@tempa \fi \ifx \@tempb \@empty \def\@tempb {arXiv}\fi \@ifundefined {mn@eprint@\@tempb}{\@tempb:\@tempc}{\expandafter \expandafter \csname mn@eprint@\@tempb\endcsname \expandafter{\@tempc}}}

\bibitem[\protect\citeauthoryear{{Alp}, {Larsson}  \& {Fransson}}{{Alp} et~al.}{2021}]{2021ApJ...916...76A}
{Alp} D.,  {Larsson} J.,   {Fransson} C.,  2021, \mn@doi [\apj] {10.3847/1538-4357/ac052d}, \href {https://ui.adsabs.harvard.edu/abs/2021ApJ...916...76A} {916, 76}

\bibitem[\protect\citeauthoryear{{Arnett}}{{Arnett}}{1969}]{1969Ap&SS...5..180A}
{Arnett} W.~D.,  1969, \mn@doi [\apss] {10.1007/BF00650291}, \href {https://ui.adsabs.harvard.edu/abs/1969Ap&SS...5..180A} {5, 180}

\bibitem[\protect\citeauthoryear{{Bassi} et~al.,}{{Bassi} et~al.}{2019}]{2019MNRAS.482.1587B}
{Bassi} T.,  et~al., 2019, \mn@doi [\mnras] {10.1093/mnras/sty2739}, \href {https://ui.adsabs.harvard.edu/abs/2019MNRAS.482.1587B} {482, 1587}

\bibitem[\protect\citeauthoryear{{Blair}, {Long}  \& {Vancura}}{{Blair} et~al.}{1991}]{1991ApJ...366..484B}
{Blair} W.~P.,  {Long} K.~S.,   {Vancura} O.,  1991, \mn@doi [\apj] {10.1086/169583}, \href {https://ui.adsabs.harvard.edu/abs/1991ApJ...366..484B} {366, 484}

\bibitem[\protect\citeauthoryear{{Boggs} et~al.,}{{Boggs} et~al.}{2015}]{2015Sci...348..670B}
{Boggs} S.~E.,  et~al., 2015, \mn@doi [Science] {10.1126/science.aaa2259}, \href {https://ui.adsabs.harvard.edu/abs/2015Sci...348..670B} {348, 670}

\bibitem[\protect\citeauthoryear{{Borkowski}, {Reynolds}, {Green}, {Hwang}, {Petre}, {Krishnamurthy}  \& {Willett}}{{Borkowski} et~al.}{2010}]{2010ApJ...724L.161B}
{Borkowski} K.~J.,  {Reynolds} S.~P.,  {Green} D.~A.,  {Hwang} U.,  {Petre} R.,  {Krishnamurthy} K.,   {Willett} R.,  2010, \mn@doi [\apjl] {10.1088/2041-8205/724/2/L161}, \href {https://ui.adsabs.harvard.edu/abs/2010ApJ...724L.161B} {724, L161}

\bibitem[\protect\citeauthoryear{{Borkowski}, {Reynolds}, {Green}, {Hwang}, {Petre}, {Krishnamurthy}  \& {Willett}}{{Borkowski} et~al.}{2013a}]{2013HEAD...1312706B}
{Borkowski} K.~J.,  {Reynolds} S.~P.,  {Green} D.,  {Hwang} U.,  {Petre} R.,  {Krishnamurthy} K.,   {Willett} R.,  2013a, in AAS/High Energy Astrophysics Division \#13. p. 127.06

\bibitem[\protect\citeauthoryear{{Borkowski}, {Reynolds}, {Hwang}, {Green}, {Petre}, {Krishnamurthy}  \& {Willett}}{{Borkowski} et~al.}{2013b}]{2013ApJ...771L...9B}
{Borkowski} K.~J.,  {Reynolds} S.~P.,  {Hwang} U.,  {Green} D.~A.,  {Petre} R.,  {Krishnamurthy} K.,   {Willett} R.,  2013b, \mn@doi [\apjl] {10.1088/2041-8205/771/1/L9}, \href {https://ui.adsabs.harvard.edu/abs/2013ApJ...771L...9B} {771, L9}

\bibitem[\protect\citeauthoryear{{Bozzetto} et~al.,}{{Bozzetto} et~al.}{2017}]{2017ApJS..230....2B}
{Bozzetto} L.~M.,  et~al., 2017, \mn@doi [\apjs] {10.3847/1538-4365/aa653c}, \href {https://ui.adsabs.harvard.edu/abs/2017ApJS..230....2B} {230, 2}

\bibitem[\protect\citeauthoryear{{Carlton}, {Borkowski}, {Reynolds}, {Hwang}, {Petre}, {Green}, {Krishnamurthy}  \& {Willett}}{{Carlton} et~al.}{2011}]{2011ApJ...737L..22C}
{Carlton} A.~K.,  {Borkowski} K.~J.,  {Reynolds} S.~P.,  {Hwang} U.,  {Petre} R.,  {Green} D.~A.,  {Krishnamurthy} K.,   {Willett} R.,  2011, \mn@doi [\apjl] {10.1088/2041-8205/737/1/L22}, \href {https://ui.adsabs.harvard.edu/abs/2011ApJ...737L..22C} {737, L22}

\bibitem[\protect\citeauthoryear{{Cash}}{{Cash}}{1979}]{1979ApJ...228..939C}
{Cash} W.,  1979, \mn@doi [\apj] {10.1086/156922}, \href {https://ui.adsabs.harvard.edu/abs/1979ApJ...228..939C} {228, 939}

\bibitem[\protect\citeauthoryear{{Chen}, {Singh}  \& {Cameron}}{{Chen} et~al.}{2011}]{2011NDS...112.2357C}
{Chen} J.,  {Singh} B.,   {Cameron} J.~A.,  2011, \mn@doi [Nuclear Data Sheets] {10.1016/j.nds.2011.08.005}, \href {https://ui.adsabs.harvard.edu/abs/2011NDS...112.2357C} {112, 2357}

\bibitem[\protect\citeauthoryear{{Chiotellis}, {Schure}  \& {Vink}}{{Chiotellis} et~al.}{2012}]{2012A&A...537A.139C}
{Chiotellis} A.,  {Schure} K.~M.,   {Vink} J.,  2012, \mn@doi [\aap] {10.1051/0004-6361/201014754}, \href {https://ui.adsabs.harvard.edu/abs/2012A&A...537A.139C} {537, A139}

\bibitem[\protect\citeauthoryear{{Diehl}}{{Diehl}}{2022}]{2022hxga.book...62D}
{Diehl} R.,  2022, in , Handbook of X-ray and Gamma-ray Astrophysics.
p.~62, \mn@doi{10.1007/978-981-16-4544-0_86-1}

\bibitem[\protect\citeauthoryear{{Dufour} \& {Kaspi}}{{Dufour} \& {Kaspi}}{2013}]{2013ApJ...775...52D}
{Dufour} F.,  {Kaspi} V.~M.,  2013, \mn@doi [\apj] {10.1088/0004-637X/775/1/52}, \href {https://ui.adsabs.harvard.edu/abs/2013ApJ...775...52D} {775, 52}

\bibitem[\protect\citeauthoryear{{Dupraz}, {Bloemen}, {Bennett}, {Diehl}, {Hermsen}, {Iyudin}, {Ryan}  \& {Schoenfelder}}{{Dupraz} et~al.}{1997}]{1997A&A...324..683D}
{Dupraz} C.,  {Bloemen} H.,  {Bennett} K.,  {Diehl} R.,  {Hermsen} W.,  {Iyudin} A.~F.,  {Ryan} J.,   {Schoenfelder} V.,  1997, \aap, \href {https://ui.adsabs.harvard.edu/abs/1997A&A...324..683D} {324, 683}

\bibitem[\protect\citeauthoryear{{Evans} et~al.,}{{Evans} et~al.}{2020}]{2020AAS...23515405E}
{Evans} I.~N.,  et~al., 2020, in American Astronomical Society Meeting Abstracts \#235. p. 154.05

\bibitem[\protect\citeauthoryear{{Ferrand} \& {Safi-Harb}}{{Ferrand} \& {Safi-Harb}}{2012}]{2012AdSpR..49.1313F}
{Ferrand} G.,  {Safi-Harb} S.,  2012, \mn@doi [Advances in Space Research] {10.1016/j.asr.2012.02.004}, \href {https://ui.adsabs.harvard.edu/abs/2012AdSpR..49.1313F} {49, 1313}

\bibitem[\protect\citeauthoryear{{Fesen}, {Hamilton}  \& {Saken}}{{Fesen} et~al.}{1989}]{1989ApJ...341L..55F}
{Fesen} R.~A.,  {Hamilton} A. J.~S.,   {Saken} J.~M.,  1989, \mn@doi [\apjl] {10.1086/185456}, \href {https://ui.adsabs.harvard.edu/abs/1989ApJ...341L..55F} {341, L55}

\bibitem[\protect\citeauthoryear{{Fesen}, {Gerardy}, {McLin}  \& {Hamilton}}{{Fesen} et~al.}{1999}]{1999ApJ...514..195F}
{Fesen} R.~A.,  {Gerardy} C.~L.,  {McLin} K.~M.,   {Hamilton} A. J.~S.,  1999, \mn@doi [\apj] {10.1086/306938}, \href {https://ui.adsabs.harvard.edu/abs/1999ApJ...514..195F} {514, 195}

\bibitem[\protect\citeauthoryear{{Fesen}, {H{\"o}flich}, {Hamilton}, {Hammell}, {Gerardy}, {Khokhlov}  \& {Wheeler}}{{Fesen} et~al.}{2007}]{2007ApJ...658..396F}
{Fesen} R.~A.,  {H{\"o}flich} P.~A.,  {Hamilton} A. J.~S.,  {Hammell} M.~C.,  {Gerardy} C.~L.,  {Khokhlov} A.~M.,   {Wheeler} J.~C.,  2007, \mn@doi [\apj] {10.1086/510998}, \href {https://ui.adsabs.harvard.edu/abs/2007ApJ...658..396F} {658, 396}

\bibitem[\protect\citeauthoryear{{Fesen}, {H{\"o}flich}  \& {Hamilton}}{{Fesen} et~al.}{2015}]{2015ApJ...804..140F}
{Fesen} R.~A.,  {H{\"o}flich} P.~A.,   {Hamilton} A. J.~S.,  2015, \mn@doi [\apj] {10.1088/0004-637X/804/2/140}, \href {https://ui.adsabs.harvard.edu/abs/2015ApJ...804..140F} {804, 140}

\bibitem[\protect\citeauthoryear{{Fesen}, {Weil}, {Hamilton}  \& {H{\"o}flich}}{{Fesen} et~al.}{2017}]{2017ApJ...848..130F}
{Fesen} R.~A.,  {Weil} K.~E.,  {Hamilton} A. J.~S.,   {H{\"o}flich} P.~A.,  2017, \mn@doi [\apj] {10.3847/1538-4357/aa8b11}, \href {https://ui.adsabs.harvard.edu/abs/2017ApJ...848..130F} {848, 130}

\bibitem[\protect\citeauthoryear{{Fink} et~al.,}{{Fink} et~al.}{2014}]{2014MNRAS.438.1762F}
{Fink} M.,  et~al., 2014, \mn@doi [\mnras] {10.1093/mnras/stt2315}, \href {https://ui.adsabs.harvard.edu/abs/2014MNRAS.438.1762F} {438, 1762}

\bibitem[\protect\citeauthoryear{{Gaia Collaboration} et~al.,}{{Gaia Collaboration} et~al.}{2021}]{2021A&A...649A...1G}
{Gaia Collaboration} et~al., 2021, \mn@doi [\aap] {10.1051/0004-6361/202039657}, \href {https://ui.adsabs.harvard.edu/abs/2021A&A...649A...1G} {649, A1}

\bibitem[\protect\citeauthoryear{{Ghavamian}, {Winkler}, {Raymond}  \& {Long}}{{Ghavamian} et~al.}{2002}]{2002ApJ...572..888G}
{Ghavamian} P.,  {Winkler} P.~F.,  {Raymond} J.~C.,   {Long} K.~S.,  2002, \mn@doi [\apj] {10.1086/340437}, \href {https://ui.adsabs.harvard.edu/abs/2002ApJ...572..888G} {572, 888}

\bibitem[\protect\citeauthoryear{{Gonz{\'a}lez Hern{\'a}ndez}, {Ruiz-Lapuente}, {Tabernero}, {Montes}, {Canal}, {M{\'e}ndez}  \& {Bedin}}{{Gonz{\'a}lez Hern{\'a}ndez} et~al.}{2012}]{2012Natur.489..533G}
{Gonz{\'a}lez Hern{\'a}ndez} J.~I.,  {Ruiz-Lapuente} P.,  {Tabernero} H.~M.,  {Montes} D.,  {Canal} R.,  {M{\'e}ndez} J.,   {Bedin} L.~R.,  2012, \mn@doi [\nat] {10.1038/nature11447}, \href {https://ui.adsabs.harvard.edu/abs/2012Natur.489..533G} {489, 533}

\bibitem[\protect\citeauthoryear{{Grefenstette} et~al.,}{{Grefenstette} et~al.}{2014}]{2014Natur.506..339G}
{Grefenstette} B.~W.,  et~al., 2014, \mn@doi [\nat] {10.1038/nature12997}, \href {https://ui.adsabs.harvard.edu/abs/2014Natur.506..339G} {506, 339}

\bibitem[\protect\citeauthoryear{{Grefenstette} et~al.,}{{Grefenstette} et~al.}{2017}]{2017ApJ...834...19G}
{Grefenstette} B.~W.,  et~al., 2017, \mn@doi [\apj] {10.3847/1538-4357/834/1/19}, \href {https://ui.adsabs.harvard.edu/abs/2017ApJ...834...19G} {834, 19}

\bibitem[\protect\citeauthoryear{{Grefenstette} et~al.,}{{Grefenstette} et~al.}{2021}]{2021ApJ...909...30G}
{Grefenstette} B.~W.,  et~al., 2021, \mn@doi [\apj] {10.3847/1538-4357/abe045}, \href {https://ui.adsabs.harvard.edu/abs/2021ApJ...909...30G} {909, 30}

\bibitem[\protect\citeauthoryear{{Gronow}, {Collins}, {Ohlmann}, {Pakmor}, {Kromer}, {Seitenzahl}, {Sim}  \& {R{\"o}pke}}{{Gronow} et~al.}{2020}]{Gronow2020}
{Gronow} S.,  {Collins} C.,  {Ohlmann} S.~T.,  {Pakmor} R.,  {Kromer} M.,  {Seitenzahl} I.~R.,  {Sim} S.~A.,   {R{\"o}pke} F.~K.,  2020, \mn@doi [\aap] {10.1051/0004-6361/201936494}, \href {https://ui.adsabs.harvard.edu/abs/2020A&A...635A.169G} {635, A169}

\bibitem[\protect\citeauthoryear{{Gronow}, {Collins}, {Sim}  \& {R{\"o}pke}}{{Gronow} et~al.}{2021a}]{Gronow2021}
{Gronow} S.,  {Collins} C.~E.,  {Sim} S.~A.,   {R{\"o}pke} F.~K.,  2021a, \mn@doi [\aap] {10.1051/0004-6361/202039954}, \href {https://ui.adsabs.harvard.edu/abs/2021A&A...649A.155G} {649, A155}

\bibitem[\protect\citeauthoryear{{Gronow}, {C{\^o}t{\'e}}, {Lach}, {Seitenzahl}, {Collins}, {Sim}  \& {R{\"o}pke}}{{Gronow} et~al.}{2021b}]{Gronow2021b}
{Gronow} S.,  {C{\^o}t{\'e}} B.,  {Lach} F.,  {Seitenzahl} I.~R.,  {Collins} C.~E.,  {Sim} S.~A.,   {R{\"o}pke} F.~K.,  2021b, \mn@doi [\aap] {10.1051/0004-6361/202140881}, \href {https://ui.adsabs.harvard.edu/abs/2021A&A...656A..94G} {656, A94}

\bibitem[\protect\citeauthoryear{{Hamilton} \& {Fesen}}{{Hamilton} \& {Fesen}}{2000}]{2000ApJ...542..779H}
{Hamilton} A. J.~S.,  {Fesen} R.~A.,  2000, \mn@doi [\apj] {10.1086/317014}, \href {https://ui.adsabs.harvard.edu/abs/2000ApJ...542..779H} {542, 779}

\bibitem[\protect\citeauthoryear{{Harp}, {Liebe}, {Craig}, {Harrison}, {Kruse-Madsen}  \& {Zoglauer}}{{Harp} et~al.}{2010}]{2010SPIE.7738E..0ZH}
{Harp} D.~I.,  {Liebe} C.~C.,  {Craig} W.,  {Harrison} F.,  {Kruse-Madsen} K.,   {Zoglauer} A.,  2010, in {Angeli} G.~Z.,  {Dierickx} P.,  eds,  Society of Photo-Optical Instrumentation Engineers (SPIE) Conference Series Vol. 7738, Modeling, Systems Engineering, and Project Management for Astronomy IV. p. 77380Z, \mn@doi{10.1117/12.856626}

\bibitem[\protect\citeauthoryear{{Harrison} et~al.,}{{Harrison} et~al.}{2013}]{2013ApJ...770..103H}
{Harrison} F.~A.,  et~al., 2013, \mn@doi [\apj] {10.1088/0004-637X/770/2/103}, \href {https://ui.adsabs.harvard.edu/abs/2013ApJ...770..103H} {770, 103}

\bibitem[\protect\citeauthoryear{{Horowitz} \& {Caplan}}{{Horowitz} \& {Caplan}}{2021}]{2021PhRvL.126m1101H}
{Horowitz} C.~J.,  {Caplan} M.~E.,  2021, \mn@doi [\prl] {10.1103/PhysRevLett.126.131101}, \href {https://ui.adsabs.harvard.edu/abs/2021PhRvL.126m1101H} {126, 131101}

\bibitem[\protect\citeauthoryear{{Iben} \& {Tutukov}}{{Iben} \& {Tutukov}}{1984}]{1984ApJS...54..335I}
{Iben} I. J.,  {Tutukov} A.~V.,  1984, \mn@doi [\apjs] {10.1086/190932}, \href {https://ui.adsabs.harvard.edu/abs/1984ApJS...54..335I} {54, 335}

\bibitem[\protect\citeauthoryear{{Iwamoto}, {Brachwitz}, {Nomoto}, {Kishimoto}, {Umeda}, {Hix}  \& {Thielemann}}{{Iwamoto} et~al.}{1999}]{1999ApJS..125..439I}
{Iwamoto} K.,  {Brachwitz} F.,  {Nomoto} K.,  {Kishimoto} N.,  {Umeda} H.,  {Hix} W.~R.,   {Thielemann} F.-K.,  1999, \mn@doi [\apjs] {10.1086/313278}, \href {https://ui.adsabs.harvard.edu/abs/1999ApJS..125..439I} {125, 439}

\bibitem[\protect\citeauthoryear{{Iyudin} et~al.,}{{Iyudin} et~al.}{1999}]{1999ApL&C..38..383I}
{Iyudin} A.~F.,  et~al., 1999, Astrophysical Letters and Communications, \href {https://ui.adsabs.harvard.edu/abs/1999ApL&C..38..383I} {38, 383}

\bibitem[\protect\citeauthoryear{{Jha}, {Maguire}  \& {Sullivan}}{{Jha} et~al.}{2019}]{2019NatAs...3..706J}
{Jha} S.~W.,  {Maguire} K.,   {Sullivan} M.,  2019, \mn@doi [Nature Astronomy] {10.1038/s41550-019-0858-0}, \href {https://ui.adsabs.harvard.edu/abs/2019NatAs...3..706J} {3, 706}

\bibitem[\protect\citeauthoryear{{Kaastra} \& {Bleeker}}{{Kaastra} \& {Bleeker}}{2016}]{2016A&A...587A.151K}
{Kaastra} J.~S.,  {Bleeker} J.~A.~M.,  2016, \mn@doi [\aap] {10.1051/0004-6361/201527395}, \href {https://ui.adsabs.harvard.edu/abs/2016A&A...587A.151K} {587, A151}

\bibitem[\protect\citeauthoryear{{Karakas}}{{Karakas}}{2010}]{2010MNRAS.403.1413K}
{Karakas} A.~I.,  2010, \mn@doi [\mnras] {10.1111/j.1365-2966.2009.16198.x}, \href {https://ui.adsabs.harvard.edu/abs/2010MNRAS.403.1413K} {403, 1413}

\bibitem[\protect\citeauthoryear{{Kashi} \& {Soker}}{{Kashi} \& {Soker}}{2011}]{2011MNRAS.417.1466K}
{Kashi} A.,  {Soker} N.,  2011, \mn@doi [\mnras] {10.1111/j.1365-2966.2011.19361.x}, \href {https://ui.adsabs.harvard.edu/abs/2011MNRAS.417.1466K} {417, 1466}

\bibitem[\protect\citeauthoryear{{Kasuga}, {Vink}, {Katsuda}, {Uchida}, {Bamba}, {Sato}  \& {Hughes}}{{Kasuga} et~al.}{2021}]{2021ApJ...915...42K}
{Kasuga} T.,  {Vink} J.,  {Katsuda} S.,  {Uchida} H.,  {Bamba} A.,  {Sato} T.,   {Hughes} J.~P.,  2021, \mn@doi [\apj] {10.3847/1538-4357/abff4f}, \href {https://ui.adsabs.harvard.edu/abs/2021ApJ...915...42K} {915, 42}

\bibitem[\protect\citeauthoryear{{Katsuda} et~al.,}{{Katsuda} et~al.}{2015}]{2015ApJ...808...49K}
{Katsuda} S.,  et~al., 2015, \mn@doi [\apj] {10.1088/0004-637X/808/1/49}, \href {https://ui.adsabs.harvard.edu/abs/2015ApJ...808...49K} {808, 49}

\bibitem[\protect\citeauthoryear{{Kerzendorf}, {Schmidt}, {Laird}, {Podsiadlowski}  \& {Bessell}}{{Kerzendorf} et~al.}{2012}]{2012ApJ...759....7K}
{Kerzendorf} W.~E.,  {Schmidt} B.~P.,  {Laird} J.~B.,  {Podsiadlowski} P.,   {Bessell} M.~S.,  2012, \mn@doi [\apj] {10.1088/0004-637X/759/1/7}, \href {https://ui.adsabs.harvard.edu/abs/2012ApJ...759....7K} {759, 7}

\bibitem[\protect\citeauthoryear{{Kerzendorf}, {Childress}, {Scharw{\"a}chter}, {Do}  \& {Schmidt}}{{Kerzendorf} et~al.}{2014}]{2014ApJ...782...27K}
{Kerzendorf} W.~E.,  {Childress} M.,  {Scharw{\"a}chter} J.,  {Do} T.,   {Schmidt} B.~P.,  2014, \mn@doi [\apj] {10.1088/0004-637X/782/1/27}, \href {https://ui.adsabs.harvard.edu/abs/2014ApJ...782...27K} {782, 27}

\bibitem[\protect\citeauthoryear{{Kerzendorf}, {Strampelli}, {Shen}, {Schwab}, {Pakmor}, {Do}, {Buchner}  \& {Rest}}{{Kerzendorf} et~al.}{2018}]{2018MNRAS.479..192K}
{Kerzendorf} W.~E.,  {Strampelli} G.,  {Shen} K.~J.,  {Schwab} J.,  {Pakmor} R.,  {Do} T.,  {Buchner} J.,   {Rest} A.,  2018, \mn@doi [\mnras] {10.1093/mnras/sty1357}, \href {https://ui.adsabs.harvard.edu/abs/2018MNRAS.479..192K} {479, 192}

\bibitem[\protect\citeauthoryear{{Khokhlov}}{{Khokhlov}}{1991}]{1991A&A...245..114K}
{Khokhlov} A.~M.,  1991, \aap, \href {https://ui.adsabs.harvard.edu/abs/1991A&A...245..114K} {245, 114}

\bibitem[\protect\citeauthoryear{{Kosakowski}, {Ugalino}, {Fisher}, {Graur}, {Bobrick}  \& {Perets}}{{Kosakowski} et~al.}{2023}]{2023MNRAS.519L..74K}
{Kosakowski} D.,  {Ugalino} M.~I.,  {Fisher} R.,  {Graur} O.,  {Bobrick} A.,   {Perets} H.~B.,  2023, \mn@doi [\mnras] {10.1093/mnrasl/slac152}, \href {https://ui.adsabs.harvard.edu/abs/2023MNRAS.519L..74K} {519, L74}

\bibitem[\protect\citeauthoryear{{Krimm} et~al.,}{{Krimm} et~al.}{2013}]{2013ApJS..209...14K}
{Krimm} H.~A.,  et~al., 2013, \mn@doi [\apjs] {10.1088/0067-0049/209/1/14}, \href {https://ui.adsabs.harvard.edu/abs/2013ApJS..209...14K} {209, 14}

\bibitem[\protect\citeauthoryear{{Kromer}, {Sim}, {Fink}, {R{\"o}pke}, {Seitenzahl}  \& {Hillebrandt}}{{Kromer} et~al.}{2010}]{2010ApJ...719.1067K}
{Kromer} M.,  {Sim} S.~A.,  {Fink} M.,  {R{\"o}pke} F.~K.,  {Seitenzahl} I.~R.,   {Hillebrandt} W.,  2010, \mn@doi [\apj] {10.1088/0004-637X/719/2/1067}, \href {https://ui.adsabs.harvard.edu/abs/2010ApJ...719.1067K} {719, 1067}

\bibitem[\protect\citeauthoryear{{Kromer} et~al.,}{{Kromer} et~al.}{2013}]{2013ApJ...778L..18K}
{Kromer} M.,  et~al., 2013, \mn@doi [\apjl] {10.1088/2041-8205/778/1/L18}, \href {https://ui.adsabs.harvard.edu/abs/2013ApJ...778L..18K} {778, L18}

\bibitem[\protect\citeauthoryear{{Kromer} et~al.,}{{Kromer} et~al.}{2015}]{2015MNRAS.450.3045K}
{Kromer} M.,  et~al., 2015, \mn@doi [\mnras] {10.1093/mnras/stv886}, \href {https://ui.adsabs.harvard.edu/abs/2015MNRAS.450.3045K} {450, 3045}

\bibitem[\protect\citeauthoryear{{Kromer} et~al.,}{{Kromer} et~al.}{2016}]{2016MNRAS.459.4428K}
{Kromer} M.,  et~al., 2016, \mn@doi [\mnras] {10.1093/mnras/stw962}, \href {https://ui.adsabs.harvard.edu/abs/2016MNRAS.459.4428K} {459, 4428}

\bibitem[\protect\citeauthoryear{{Kushnir}, {Katz}, {Dong}, {Livne}  \& {Fern{\'a}ndez}}{{Kushnir} et~al.}{2013}]{2013ApJ...778L..37K}
{Kushnir} D.,  {Katz} B.,  {Dong} S.,  {Livne} E.,   {Fern{\'a}ndez} R.,  2013, \mn@doi [\apjl] {10.1088/2041-8205/778/2/L37}, \href {https://ui.adsabs.harvard.edu/abs/2013ApJ...778L..37K} {778, L37}

\bibitem[\protect\citeauthoryear{{Lach}, {R{\"o}pke}, {Seitenzahl}, {Cot{\'e}}, {Gronow}  \& {Ruiter}}{{Lach} et~al.}{2020}]{2020A&A...644A.118L}
{Lach} F.,  {R{\"o}pke} F.~K.,  {Seitenzahl} I.~R.,  {Cot{\'e}} B.,  {Gronow} S.,   {Ruiter} A.~J.,  2020, \mn@doi [\aap] {10.1051/0004-6361/202038721}, \href {https://ui.adsabs.harvard.edu/abs/2020A&A...644A.118L} {644, A118}

\bibitem[\protect\citeauthoryear{{Lach}, {Callan}, {Bubeck}, {R{\"o}pke}, {Sim}, {Schrauth}, {Ohlmann}  \& {Kromer}}{{Lach} et~al.}{2022a}]{Lach2022}
{Lach} F.,  {Callan} F.~P.,  {Bubeck} D.,  {R{\"o}pke} F.~K.,  {Sim} S.~A.,  {Schrauth} M.,  {Ohlmann} S.~T.,   {Kromer} M.,  2022a, \mn@doi [\aap] {10.1051/0004-6361/202141453}, \href {https://ui.adsabs.harvard.edu/abs/2022A&A...658A.179L} {658, A179}

\bibitem[\protect\citeauthoryear{{Lach}, {Callan}, {Sim}  \& {R{\"o}pke}}{{Lach} et~al.}{2022b}]{Lach2022b}
{Lach} F.,  {Callan} F.~P.,  {Sim} S.~A.,   {R{\"o}pke} F.~K.,  2022b, \mn@doi [\aap] {10.1051/0004-6361/202142194}, \href {https://ui.adsabs.harvard.edu/abs/2022A&A...659A..27L} {659, A27}

\bibitem[\protect\citeauthoryear{{Leung} \& {Nomoto}}{{Leung} \& {Nomoto}}{2018}]{2018ApJ...861..143L}
{Leung} S.-C.,  {Nomoto} K.,  2018, \mn@doi [\apj] {10.3847/1538-4357/aac2df}, \href {https://ui.adsabs.harvard.edu/abs/2018ApJ...861..143L} {861, 143}

\bibitem[\protect\citeauthoryear{{Leung} \& {Nomoto}}{{Leung} \& {Nomoto}}{2020a}]{2020ApJ...888...80L}
{Leung} S.-C.,  {Nomoto} K.,  2020a, \mn@doi [\apj] {10.3847/1538-4357/ab5c1f}, \href {https://ui.adsabs.harvard.edu/abs/2020ApJ...888...80L} {888, 80}

\bibitem[\protect\citeauthoryear{{Leung} \& {Nomoto}}{{Leung} \& {Nomoto}}{2020b}]{2020ApJ...900...54L}
{Leung} S.-C.,  {Nomoto} K.,  2020b, \mn@doi [\apj] {10.3847/1538-4357/aba1e3}, \href {https://ui.adsabs.harvard.edu/abs/2020ApJ...900...54L} {900, 54}

\bibitem[\protect\citeauthoryear{{Li} et~al.,}{{Li} et~al.}{2018}]{2018ApJ...864...85L}
{Li} J.-T.,  et~al., 2018, \mn@doi [\apj] {10.3847/1538-4357/aad598}, \href {https://ui.adsabs.harvard.edu/abs/2018ApJ...864...85L} {864, 85}

\bibitem[\protect\citeauthoryear{{Livne}}{{Livne}}{1990}]{1990ApJ...354L..53L}
{Livne} E.,  1990, \mn@doi [\apjl] {10.1086/185721}, \href {https://ui.adsabs.harvard.edu/abs/1990ApJ...354L..53L} {354, L53}

\bibitem[\protect\citeauthoryear{{Luken} et~al.,}{{Luken} et~al.}{2020}]{2020MNRAS.492.2606L}
{Luken} K.~J.,  et~al., 2020, \mn@doi [\mnras] {10.1093/mnras/stz3439}, \href {https://ui.adsabs.harvard.edu/abs/2020MNRAS.492.2606L} {492, 2606}

\bibitem[\protect\citeauthoryear{{Madsen}, {Christensen}, {Craig}, {Forster}, {Grefenstette}, {Harrison}, {Miyasaka}  \& {Rana}}{{Madsen} et~al.}{2017a}]{2017JATIS...3d4003M}
{Madsen} K.~K.,  {Christensen} F.~E.,  {Craig} W.~W.,  {Forster} K.~W.,  {Grefenstette} B.~W.,  {Harrison} F.~A.,  {Miyasaka} H.,   {Rana} V.,  2017a, \mn@doi [Journal of Astronomical Telescopes, Instruments, and Systems] {10.1117/1.JATIS.3.4.044003}, \href {https://ui.adsabs.harvard.edu/abs/2017JATIS...3d4003M} {3, 044003}

\bibitem[\protect\citeauthoryear{{Madsen}, {Forster}, {Grefenstette}, {Harrison}  \& {Stern}}{{Madsen} et~al.}{2017b}]{2017ApJ...841...56M}
{Madsen} K.~K.,  {Forster} K.,  {Grefenstette} B.~W.,  {Harrison} F.~A.,   {Stern} D.,  2017b, \mn@doi [\apj] {10.3847/1538-4357/aa6970}, \href {https://ui.adsabs.harvard.edu/abs/2017ApJ...841...56M} {841, 56}

\bibitem[\protect\citeauthoryear{{Maeda}, {R{\"o}pke}, {Fink}, {Hillebrandt}, {Travaglio}  \& {Thielemann}}{{Maeda} et~al.}{2010}]{2010ApJ...712..624M}
{Maeda} K.,  {R{\"o}pke} F.~K.,  {Fink} M.,  {Hillebrandt} W.,  {Travaglio} C.,   {Thielemann} F.~K.,  2010, \mn@doi [\apj] {10.1088/0004-637X/712/1/624}, \href {https://ui.adsabs.harvard.edu/abs/2010ApJ...712..624M} {712, 624}

\bibitem[\protect\citeauthoryear{{Maggi} et~al.,}{{Maggi} et~al.}{2016}]{2016A&A...585A.162M}
{Maggi} P.,  et~al., 2016, \mn@doi [\aap] {10.1051/0004-6361/201526932}, \href {https://ui.adsabs.harvard.edu/abs/2016A&A...585A.162M} {585, A162}

\bibitem[\protect\citeauthoryear{{Maggi} et~al.,}{{Maggi} et~al.}{2019}]{2019A&A...631A.127M}
{Maggi} P.,  et~al., 2019, \mn@doi [\aap] {10.1051/0004-6361/201936583}, \href {https://ui.adsabs.harvard.edu/abs/2019A&A...631A.127M} {631, A127}

\bibitem[\protect\citeauthoryear{{Marquardt}, {Sim}, {Ruiter}, {Seitenzahl}, {Ohlmann}, {Kromer}, {Pakmor}  \& {R{\"o}pke}}{{Marquardt} et~al.}{2015}]{Marquardt2015}
{Marquardt} K.~S.,  {Sim} S.~A.,  {Ruiter} A.~J.,  {Seitenzahl} I.~R.,  {Ohlmann} S.~T.,  {Kromer} M.,  {Pakmor} R.,   {R{\"o}pke} F.~K.,  2015, \mn@doi [\aap] {10.1051/0004-6361/201525761}, \href {https://ui.adsabs.harvard.edu/abs/2015A&A...580A.118M} {580, A118}

\bibitem[\protect\citeauthoryear{{McConnachie}}{{McConnachie}}{2012}]{2012AJ....144....4M}
{McConnachie} A.~W.,  2012, \mn@doi [\aj] {10.1088/0004-6256/144/1/4}, \href {https://ui.adsabs.harvard.edu/abs/2012AJ....144....4M} {144, 4}

\bibitem[\protect\citeauthoryear{{Mernier} et~al.,}{{Mernier} et~al.}{2020}]{2020AN....341..203M}
{Mernier} F.,  et~al., 2020, \mn@doi [Astronomische Nachrichten] {10.1002/asna.202023779}, \href {https://ui.adsabs.harvard.edu/abs/2020AN....341..203M} {341, 203}

\bibitem[\protect\citeauthoryear{{Mewe}}{{Mewe}}{1999}]{1999LNP...520..109M}
{Mewe} R.,  1999, in {van Paradijs} J.,  {Bleeker} J. A.~M.,  eds, , Vol.~520, X-Ray Spectroscopy in Astrophysics.
p.~109, \mn@doi{10.1007/978-3-540-49199-6_2}

\bibitem[\protect\citeauthoryear{{Moll} \& {Woosley}}{{Moll} \& {Woosley}}{2013}]{2013ApJ...774..137M}
{Moll} R.,  {Woosley} S.~E.,  2013, \mn@doi [\apj] {10.1088/0004-637X/774/2/137}, \href {https://ui.adsabs.harvard.edu/abs/2013ApJ...774..137M} {774, 137}

\bibitem[\protect\citeauthoryear{{Neopane}, {Bhargava}, {Fisher}, {Ferrari}, {Yoshida}, {Toonen}  \& {Bravo}}{{Neopane} et~al.}{2022}]{2022ApJ...925...92N}
{Neopane} S.,  {Bhargava} K.,  {Fisher} R.,  {Ferrari} M.,  {Yoshida} S.,  {Toonen} S.,   {Bravo} E.,  2022, \mn@doi [\apj] {10.3847/1538-4357/ac3b52}, \href {https://ui.adsabs.harvard.edu/abs/2022ApJ...925...92N} {925, 92}

\bibitem[\protect\citeauthoryear{{Nikoli{\'c}}, {van de Ven}, {Heng}, {Kupko}, {Husemann}, {Raymond}, {Hughes}  \& {Falc{\'o}n-Barroso}}{{Nikoli{\'c}} et~al.}{2013}]{2013Sci...340...45N}
{Nikoli{\'c}} S.,  {van de Ven} G.,  {Heng} K.,  {Kupko} D.,  {Husemann} B.,  {Raymond} J.~C.,  {Hughes} J.~P.,   {Falc{\'o}n-Barroso} J.,  2013, \mn@doi [Science] {10.1126/science.1228297}, \href {https://ui.adsabs.harvard.edu/abs/2013Sci...340...45N} {340, 45}

\bibitem[\protect\citeauthoryear{{Nomoto}}{{Nomoto}}{1982}]{1982ApJ...253..798N}
{Nomoto} K.,  1982, \mn@doi [\apj] {10.1086/159682}, \href {https://ui.adsabs.harvard.edu/abs/1982ApJ...253..798N} {253, 798}

\bibitem[\protect\citeauthoryear{{Nomoto}, {Thielemann}  \& {Yokoi}}{{Nomoto} et~al.}{1984}]{1984ApJ...286..644N}
{Nomoto} K.,  {Thielemann} F.~K.,   {Yokoi} K.,  1984, \mn@doi [\apj] {10.1086/162639}, \href {https://ui.adsabs.harvard.edu/abs/1984ApJ...286..644N} {286, 644}

\bibitem[\protect\citeauthoryear{{Ohlmann}, {Kromer}, {Fink}, {Pakmor}, {Seitenzahl}, {Sim}  \& {R{\"o}pke}}{{Ohlmann} et~al.}{2014}]{Ohlmann2014}
{Ohlmann} S.~T.,  {Kromer} M.,  {Fink} M.,  {Pakmor} R.,  {Seitenzahl} I.~R.,  {Sim} S.~A.,   {R{\"o}pke} F.~K.,  2014, \mn@doi [\aap] {10.1051/0004-6361/201423924}, \href {https://ui.adsabs.harvard.edu/abs/2014A&A...572A..57O} {572, A57}

\bibitem[\protect\citeauthoryear{{Pakmor}, {Kromer}, {R{\"o}pke}, {Sim}, {Ruiter}  \& {Hillebrandt}}{{Pakmor} et~al.}{2010}]{2010Natur.463...61P}
{Pakmor} R.,  {Kromer} M.,  {R{\"o}pke} F.~K.,  {Sim} S.~A.,  {Ruiter} A.~J.,   {Hillebrandt} W.,  2010, \mn@doi [\nat] {10.1038/nature08642}, \href {https://ui.adsabs.harvard.edu/abs/2010Natur.463...61P} {463, 61}

\bibitem[\protect\citeauthoryear{{Pakmor}, {Kromer}, {Taubenberger}, {Sim}, {R{\"o}pke}  \& {Hillebrandt}}{{Pakmor} et~al.}{2012}]{2012ApJ...747L..10P}
{Pakmor} R.,  {Kromer} M.,  {Taubenberger} S.,  {Sim} S.~A.,  {R{\"o}pke} F.~K.,   {Hillebrandt} W.,  2012, \mn@doi [\apjl] {10.1088/2041-8205/747/1/L10}, \href {https://ui.adsabs.harvard.edu/abs/2012ApJ...747L..10P} {747, L10}

\bibitem[\protect\citeauthoryear{{Pakmor}, {Kromer}, {Taubenberger}  \& {Springel}}{{Pakmor} et~al.}{2013}]{2013ApJ...770L...8P}
{Pakmor} R.,  {Kromer} M.,  {Taubenberger} S.,   {Springel} V.,  2013, \mn@doi [\apjl] {10.1088/2041-8205/770/1/L8}, \href {https://ui.adsabs.harvard.edu/abs/2013ApJ...770L...8P} {770, L8}

\bibitem[\protect\citeauthoryear{{Pakmor}, {Zenati}, {Perets}  \& {Toonen}}{{Pakmor} et~al.}{2021}]{2021MNRAS.503.4734P}
{Pakmor} R.,  {Zenati} Y.,  {Perets} H.~B.,   {Toonen} S.,  2021, \mn@doi [\mnras] {10.1093/mnras/stab686}, \href {https://ui.adsabs.harvard.edu/abs/2021MNRAS.503.4734P} {503, 4734}

\bibitem[\protect\citeauthoryear{{Pakmor} et~al.,}{{Pakmor} et~al.}{2022}]{2022MNRAS.517.5260P}
{Pakmor} R.,  et~al., 2022, \mn@doi [\mnras] {10.1093/mnras/stac3107}, \href {https://ui.adsabs.harvard.edu/abs/2022MNRAS.517.5260P} {517, 5260}

\bibitem[\protect\citeauthoryear{{Papish} \& {Perets}}{{Papish} \& {Perets}}{2016}]{2016ApJ...822...19P}
{Papish} O.,  {Perets} H.~B.,  2016, \mn@doi [\apj] {10.3847/0004-637X/822/1/19}, \href {https://ui.adsabs.harvard.edu/abs/2016ApJ...822...19P} {822, 19}

\bibitem[\protect\citeauthoryear{{Patnaude}, {Badenes}, {Park}  \& {Laming}}{{Patnaude} et~al.}{2012}]{2012ApJ...756....6P}
{Patnaude} D.~J.,  {Badenes} C.,  {Park} S.,   {Laming} J.~M.,  2012, \mn@doi [\apj] {10.1088/0004-637X/756/1/6}, \href {https://ui.adsabs.harvard.edu/abs/2012ApJ...756....6P} {756, 6}

\bibitem[\protect\citeauthoryear{{Perets}}{{Perets}}{2014}]{2014arXiv1407.2254P}
{Perets} H.~B.,  2014, \mn@doi [arXiv e-prints] {10.48550/arXiv.1407.2254}, \href {https://ui.adsabs.harvard.edu/abs/2014arXiv1407.2254P} {p. arXiv:1407.2254}

\bibitem[\protect\citeauthoryear{{Perets} et~al.,}{{Perets} et~al.}{2010}]{2010Natur.465..322P}
{Perets} H.~B.,  et~al., 2010, \mn@doi [\nat] {10.1038/nature09056}, \href {https://ui.adsabs.harvard.edu/abs/2010Natur.465..322P} {465, 322}

\bibitem[\protect\citeauthoryear{{Perets}, {Badenes}, {Arcavi}, {Simon}  \& {Gal-yam}}{{Perets} et~al.}{2011}]{2011ApJ...730...89P}
{Perets} H.~B.,  {Badenes} C.,  {Arcavi} I.,  {Simon} J.~D.,   {Gal-yam} A.,  2011, \mn@doi [\apj] {10.1088/0004-637X/730/2/89}, \href {https://ui.adsabs.harvard.edu/abs/2011ApJ...730...89P} {730, 89}

\bibitem[\protect\citeauthoryear{{Plewa}, {Calder}  \& {Lamb}}{{Plewa} et~al.}{2004}]{2004ApJ...612L..37P}
{Plewa} T.,  {Calder} A.~C.,   {Lamb} D.~Q.,  2004, \mn@doi [\apjl] {10.1086/424036}, \href {https://ui.adsabs.harvard.edu/abs/2004ApJ...612L..37P} {612, L37}

\bibitem[\protect\citeauthoryear{{Reid} et~al.,}{{Reid} et~al.}{2014}]{2014ApJ...783..130R}
{Reid} M.~J.,  et~al., 2014, \mn@doi [\apj] {10.1088/0004-637X/783/2/130}, \href {https://ui.adsabs.harvard.edu/abs/2014ApJ...783..130R} {783, 130}

\bibitem[\protect\citeauthoryear{{Reynolds}, {Borkowski}, {Hwang}, {Hughes}, {Badenes}, {Laming}  \& {Blondin}}{{Reynolds} et~al.}{2007}]{2007ApJ...668L.135R}
{Reynolds} S.~P.,  {Borkowski} K.~J.,  {Hwang} U.,  {Hughes} J.~P.,  {Badenes} C.,  {Laming} J.~M.,   {Blondin} J.~M.,  2007, \mn@doi [\apjl] {10.1086/522830}, \href {https://ui.adsabs.harvard.edu/abs/2007ApJ...668L.135R} {668, L135}

\bibitem[\protect\citeauthoryear{{Reynolds}, {Borkowski}, {Green}, {Hwang}, {Harrus}  \& {Petre}}{{Reynolds} et~al.}{2008}]{2008ApJ...680L..41R}
{Reynolds} S.~P.,  {Borkowski} K.~J.,  {Green} D.~A.,  {Hwang} U.,  {Harrus} I.,   {Petre} R.,  2008, \mn@doi [\apjl] {10.1086/589570}, \href {https://ui.adsabs.harvard.edu/abs/2008ApJ...680L..41R} {680, L41}

\bibitem[\protect\citeauthoryear{{Roy} et~al.,}{{Roy} et~al.}{2022}]{2022ApJ...932L..24R}
{Roy} N.~C.,  et~al., 2022, \mn@doi [\apjl] {10.3847/2041-8213/ac75e7}, \href {https://ui.adsabs.harvard.edu/abs/2022ApJ...932L..24R} {932, L24}

\bibitem[\protect\citeauthoryear{{Sankrit}, {Raymond}, {Blair}, {Long}, {Williams}, {Borkowski}, {Patnaude}  \& {Reynolds}}{{Sankrit} et~al.}{2016}]{2016ApJ...817...36S}
{Sankrit} R.,  {Raymond} J.~C.,  {Blair} W.~P.,  {Long} K.~S.,  {Williams} B.~J.,  {Borkowski} K.~J.,  {Patnaude} D.~J.,   {Reynolds} S.~P.,  2016, \mn@doi [\apj] {10.3847/0004-637X/817/1/36}, \href {https://ui.adsabs.harvard.edu/abs/2016ApJ...817...36S} {817, 36}

\bibitem[\protect\citeauthoryear{{Sano}, {Yamaguchi}, {Aruga}, {Fukui}, {Tachihara}, {Filipovi{\'c}}  \& {Rowell}}{{Sano} et~al.}{2022}]{2022ApJ...933..157S}
{Sano} H.,  {Yamaguchi} H.,  {Aruga} M.,  {Fukui} Y.,  {Tachihara} K.,  {Filipovi{\'c}} M.~D.,   {Rowell} G.,  2022, \mn@doi [\apj] {10.3847/1538-4357/ac7465}, \href {https://ui.adsabs.harvard.edu/abs/2022ApJ...933..157S} {933, 157}

\bibitem[\protect\citeauthoryear{{Sarbadhicary}, {Chomiuk}, {Badenes}, {Tremou}, {Soderberg}  \& {Sjouwerman}}{{Sarbadhicary} et~al.}{2019}]{2019ApJ...872..191S}
{Sarbadhicary} S.~K.,  {Chomiuk} L.,  {Badenes} C.,  {Tremou} E.,  {Soderberg} A.~M.,   {Sjouwerman} L.~O.,  2019, \mn@doi [\apj] {10.3847/1538-4357/ab027f}, \href {https://ui.adsabs.harvard.edu/abs/2019ApJ...872..191S} {872, 191}

\bibitem[\protect\citeauthoryear{{Seitenzahl} et~al.,}{{Seitenzahl} et~al.}{2013}]{2013MNRAS.429.1156S}
{Seitenzahl} I.~R.,  et~al., 2013, \mn@doi [\mnras] {10.1093/mnras/sts402}, \href {https://ui.adsabs.harvard.edu/abs/2013MNRAS.429.1156S} {429, 1156}

\bibitem[\protect\citeauthoryear{{Seitenzahl} et~al.,}{{Seitenzahl} et~al.}{2016}]{Seitenzahl2016}
{Seitenzahl} I.~R.,  et~al., 2016, \mn@doi [\aap] {10.1051/0004-6361/201527251}, \href {https://ui.adsabs.harvard.edu/abs/2016A&A...592A..57S} {592, A57}

\bibitem[\protect\citeauthoryear{{Shen} \& {Bildsten}}{{Shen} \& {Bildsten}}{2009}]{2009ApJ...699.1365S}
{Shen} K.~J.,  {Bildsten} L.,  2009, \mn@doi [\apj] {10.1088/0004-637X/699/2/1365}, \href {https://ui.adsabs.harvard.edu/abs/2009ApJ...699.1365S} {699, 1365}

\bibitem[\protect\citeauthoryear{{Shen} \& {Bildsten}}{{Shen} \& {Bildsten}}{2014}]{2014ApJ...785...61S}
{Shen} K.~J.,  {Bildsten} L.,  2014, \mn@doi [\apj] {10.1088/0004-637X/785/1/61}, \href {https://ui.adsabs.harvard.edu/abs/2014ApJ...785...61S} {785, 61}

\bibitem[\protect\citeauthoryear{{Shields} et~al.,}{{Shields} et~al.}{2022}]{2022ApJ...933L..31S}
{Shields} J.~V.,  et~al., 2022, \mn@doi [\apjl] {10.3847/2041-8213/ac7950}, \href {https://ui.adsabs.harvard.edu/abs/2022ApJ...933L..31S} {933, L31}

\bibitem[\protect\citeauthoryear{{Sim}, {R{\"o}pke}, {Hillebrandt}, {Kromer}, {Pakmor}, {Fink}, {Ruiter}  \& {Seitenzahl}}{{Sim} et~al.}{2010}]{2010ApJ...714L..52S}
{Sim} S.~A.,  {R{\"o}pke} F.~K.,  {Hillebrandt} W.,  {Kromer} M.,  {Pakmor} R.,  {Fink} M.,  {Ruiter} A.~J.,   {Seitenzahl} I.~R.,  2010, \mn@doi [\apjl] {10.1088/2041-8205/714/1/L52}, \href {https://ui.adsabs.harvard.edu/abs/2010ApJ...714L..52S} {714, L52}

\bibitem[\protect\citeauthoryear{{Sim}, {Fink}, {Kromer}, {R{\"o}pke}, {Ruiter}  \& {Hillebrandt}}{{Sim} et~al.}{2012}]{2012MNRAS.420.3003S}
{Sim} S.~A.,  {Fink} M.,  {Kromer} M.,  {R{\"o}pke} F.~K.,  {Ruiter} A.~J.,   {Hillebrandt} W.,  2012, \mn@doi [\mnras] {10.1111/j.1365-2966.2011.20162.x}, \href {https://ui.adsabs.harvard.edu/abs/2012MNRAS.420.3003S} {420, 3003}

\bibitem[\protect\citeauthoryear{{Soker}, {Kashi}, {Garc{\'\i}a-Berro}, {Torres}  \& {Camacho}}{{Soker} et~al.}{2013}]{2013MNRAS.431.1541S}
{Soker} N.,  {Kashi} A.,  {Garc{\'\i}a-Berro} E.,  {Torres} S.,   {Camacho} J.,  2013, \mn@doi [\mnras] {10.1093/mnras/stt271}, \href {https://ui.adsabs.harvard.edu/abs/2013MNRAS.431.1541S} {431, 1541}

\bibitem[\protect\citeauthoryear{{Sun} \& {Chen}}{{Sun} \& {Chen}}{2019}]{2019ApJ...872...45S}
{Sun} L.,  {Chen} Y.,  2019, \mn@doi [\apj] {10.3847/1538-4357/aafb73}, \href {https://ui.adsabs.harvard.edu/abs/2019ApJ...872...45S} {872, 45}

\bibitem[\protect\citeauthoryear{{The} et~al.,}{{The} et~al.}{2006}]{2006A&A...450.1037T}
{The} L.~S.,  et~al., 2006, \mn@doi [\aap] {10.1051/0004-6361:20054626}, \href {https://ui.adsabs.harvard.edu/abs/2006A&A...450.1037T} {450, 1037}

\bibitem[\protect\citeauthoryear{{Timmes}, {Woosley}, {Hartmann}  \& {Hoffman}}{{Timmes} et~al.}{1996}]{1996ApJ...464..332T}
{Timmes} F.~X.,  {Woosley} S.~E.,  {Hartmann} D.~H.,   {Hoffman} R.~D.,  1996, \mn@doi [\apj] {10.1086/177323}, \href {https://ui.adsabs.harvard.edu/abs/1996ApJ...464..332T} {464, 332}

\bibitem[\protect\citeauthoryear{{Townsley}, {Miles}, {Shen}  \& {Kasen}}{{Townsley} et~al.}{2019}]{2019ApJ...878L..38T}
{Townsley} D.~M.,  {Miles} B.~J.,  {Shen} K.~J.,   {Kasen} D.,  2019, \mn@doi [\apjl] {10.3847/2041-8213/ab27cd}, \href {https://ui.adsabs.harvard.edu/abs/2019ApJ...878L..38T} {878, L38}

\bibitem[\protect\citeauthoryear{{Travaglio}, {Hillebrandt}, {Reinecke}  \& {Thielemann}}{{Travaglio} et~al.}{2004}]{Travaglio2004}
{Travaglio} C.,  {Hillebrandt} W.,  {Reinecke} M.,   {Thielemann} F.~K.,  2004, \mn@doi [\aap] {10.1051/0004-6361:20041108}, \href {https://ui.adsabs.harvard.edu/abs/2004A&A...425.1029T} {425, 1029}

\bibitem[\protect\citeauthoryear{{Travaglio}, {Hillebrandt}  \& {Reinecke}}{{Travaglio} et~al.}{2005}]{Travaglio2005}
{Travaglio} C.,  {Hillebrandt} W.,   {Reinecke} M.,  2005, \mn@doi [\aap] {10.1051/0004-6361:20052883}, \href {https://ui.adsabs.harvard.edu/abs/2005A&A...443.1007T} {443, 1007}

\bibitem[\protect\citeauthoryear{{Troja} et~al.,}{{Troja} et~al.}{2014}]{2014ApJ...797L...6T}
{Troja} E.,  et~al., 2014, \mn@doi [\apjl] {10.1088/2041-8205/797/1/L6}, \href {https://ui.adsabs.harvard.edu/abs/2014ApJ...797L...6T} {797, L6}

\bibitem[\protect\citeauthoryear{{Tsebrenko} \& {Soker}}{{Tsebrenko} \& {Soker}}{2013}]{2013MNRAS.435..320T}
{Tsebrenko} D.,  {Soker} N.,  2013, \mn@doi [\mnras] {10.1093/mnras/stt1301}, \href {https://ui.adsabs.harvard.edu/abs/2013MNRAS.435..320T} {435, 320}

\bibitem[\protect\citeauthoryear{{Tsygankov}, {Krivonos}, {Lutovinov}, {Revnivtsev}, {Churazov}, {Sunyaev}  \& {Grebenev}}{{Tsygankov} et~al.}{2016}]{2016MNRAS.458.3411T}
{Tsygankov} S.~S.,  {Krivonos} R.~A.,  {Lutovinov} A.~A.,  {Revnivtsev} M.~G.,  {Churazov} E.~M.,  {Sunyaev} R.~A.,   {Grebenev} S.~A.,  2016, \mn@doi [\mnras] {10.1093/mnras/stw549}, \href {https://ui.adsabs.harvard.edu/abs/2016MNRAS.458.3411T} {458, 3411}

\bibitem[\protect\citeauthoryear{{Uchida}, {Yamaguchi}  \& {Koyama}}{{Uchida} et~al.}{2013}]{2013ApJ...771...56U}
{Uchida} H.,  {Yamaguchi} H.,   {Koyama} K.,  2013, \mn@doi [\apj] {10.1088/0004-637X/771/1/56}, \href {https://ui.adsabs.harvard.edu/abs/2013ApJ...771...56U} {771, 56}

\bibitem[\protect\citeauthoryear{{Vink}}{{Vink}}{2017}]{2017hsn..book.2063V}
{Vink} J.,  2017, in {Alsabti} A.~W.,  {Murdin} P.,  eds, , Handbook of Supernovae.
p.~2063, \mn@doi{10.1007/978-3-319-21846-5_92}

\bibitem[\protect\citeauthoryear{{Wagner-Kaiser}, {Sarajedini}, {Dalcanton}, {Williams}  \& {Dolphin}}{{Wagner-Kaiser} et~al.}{2015}]{2015MNRAS.451..724W}
{Wagner-Kaiser} R.,  {Sarajedini} A.,  {Dalcanton} J.~J.,  {Williams} B.~F.,   {Dolphin} A.,  2015, \mn@doi [\mnras] {10.1093/mnras/stv880}, \href {https://ui.adsabs.harvard.edu/abs/2015MNRAS.451..724W} {451, 724}

\bibitem[\protect\citeauthoryear{{Waldman}, {Sauer}, {Livne}, {Perets}, {Glasner}, {Mazzali}, {Truran}  \& {Gal-Yam}}{{Waldman} et~al.}{2011}]{2011ApJ...738...21W}
{Waldman} R.,  {Sauer} D.,  {Livne} E.,  {Perets} H.,  {Glasner} A.,  {Mazzali} P.,  {Truran} J.~W.,   {Gal-Yam} A.,  2011, \mn@doi [\apj] {10.1088/0004-637X/738/1/21}, \href {https://ui.adsabs.harvard.edu/abs/2011ApJ...738...21W} {738, 21}

\bibitem[\protect\citeauthoryear{{Webbink}}{{Webbink}}{1984}]{1984ApJ...277..355W}
{Webbink} R.~F.,  1984, \mn@doi [\apj] {10.1086/161701}, \href {https://ui.adsabs.harvard.edu/abs/1984ApJ...277..355W} {277, 355}

\bibitem[\protect\citeauthoryear{{Weinberger}, {Diehl}, {Pleintinger}, {Siegert}  \& {Greiner}}{{Weinberger} et~al.}{2020}]{2020A&A...638A..83W}
{Weinberger} C.,  {Diehl} R.,  {Pleintinger} M. M.~M.,  {Siegert} T.,   {Greiner} J.,  2020, \mn@doi [\aap] {10.1051/0004-6361/202037536}, \href {https://ui.adsabs.harvard.edu/abs/2020A&A...638A..83W} {638, A83}

\bibitem[\protect\citeauthoryear{{Weng} et~al.,}{{Weng} et~al.}{2022}]{2022ApJ...924..119W}
{Weng} J.,  et~al., 2022, \mn@doi [\apj] {10.3847/1538-4357/ac308d}, \href {https://ui.adsabs.harvard.edu/abs/2022ApJ...924..119W} {924, 119}

\bibitem[\protect\citeauthoryear{{Wenger}, {Balser}, {Anderson}  \& {Bania}}{{Wenger} et~al.}{2018}]{2018ApJ...856...52W}
{Wenger} T.~V.,  {Balser} D.~S.,  {Anderson} L.~D.,   {Bania} T.~M.,  2018, \mn@doi [\apj] {10.3847/1538-4357/aaaec8}, \href {https://ui.adsabs.harvard.edu/abs/2018ApJ...856...52W} {856, 52}

\bibitem[\protect\citeauthoryear{{Whelan} \& {Iben}}{{Whelan} \& {Iben}}{1973}]{1973ApJ...186.1007W}
{Whelan} J.,  {Iben} Icko J.,  1973, \mn@doi [\apj] {10.1086/152565}, \href {https://ui.adsabs.harvard.edu/abs/1973ApJ...186.1007W} {186, 1007}

\bibitem[\protect\citeauthoryear{{Wik} et~al.,}{{Wik} et~al.}{2014}]{2014ApJ...792...48W}
{Wik} D.~R.,  et~al., 2014, \mn@doi [\apj] {10.1088/0004-637X/792/1/48}, \href {https://ui.adsabs.harvard.edu/abs/2014ApJ...792...48W} {792, 48}

\bibitem[\protect\citeauthoryear{{Woosley} \& {Kasen}}{{Woosley} \& {Kasen}}{2011}]{2011ApJ...734...38W}
{Woosley} S.~E.,  {Kasen} D.,  2011, \mn@doi [\apj] {10.1088/0004-637X/734/1/38}, \href {https://ui.adsabs.harvard.edu/abs/2011ApJ...734...38W} {734, 38}

\bibitem[\protect\citeauthoryear{{Woosley} \& {Weaver}}{{Woosley} \& {Weaver}}{1994}]{1994ApJ...423..371W}
{Woosley} S.~E.,  {Weaver} T.~A.,  1994, \mn@doi [\apj] {10.1086/173813}, \href {https://ui.adsabs.harvard.edu/abs/1994ApJ...423..371W} {423, 371}

\bibitem[\protect\citeauthoryear{{Woosley}, {Arnett}  \& {Clayton}}{{Woosley} et~al.}{1973}]{1973ApJS...26..231W}
{Woosley} S.~E.,  {Arnett} W.~D.,   {Clayton} D.~D.,  1973, \mn@doi [\apjs] {10.1086/190282}, \href {https://ui.adsabs.harvard.edu/abs/1973ApJS...26..231W} {26, 231}

\bibitem[\protect\citeauthoryear{{Yamaguchi} et~al.,}{{Yamaguchi} et~al.}{2008}]{2008PASJ...60S.141Y}
{Yamaguchi} H.,  et~al., 2008, \mn@doi [\pasj] {10.1093/pasj/60.sp1.S141}, \href {https://ui.adsabs.harvard.edu/abs/2008PASJ...60S.141Y} {60, S141}

\bibitem[\protect\citeauthoryear{{Yukita} et~al.,}{{Yukita} et~al.}{2017}]{2017ApJ...838...47Y}
{Yukita} M.,  et~al., 2017, \mn@doi [\apj] {10.3847/1538-4357/aa62a3}, \href {https://ui.adsabs.harvard.edu/abs/2017ApJ...838...47Y} {838, 47}

\bibitem[\protect\citeauthoryear{{Zenati}, {Perets}, {Dessart}, {Jacobson-Gal{\'a}n}, {Toonen}  \& {Rest}}{{Zenati} et~al.}{2023}]{2023ApJ...944...22Z}
{Zenati} Y.,  {Perets} H.~B.,  {Dessart} L.,  {Jacobson-Gal{\'a}n} W.~V.,  {Toonen} S.,   {Rest} A.,  2023, \mn@doi [\apj] {10.3847/1538-4357/acaf65}, \href {https://ui.adsabs.harvard.edu/abs/2023ApJ...944...22Z} {944, 22}

\bibitem[\protect\citeauthoryear{{Zhekov}, {McCray}, {Borkowski}, {Burrows}  \& {Park}}{{Zhekov} et~al.}{2006}]{2006ApJ...645..293Z}
{Zhekov} S.~A.,  {McCray} R.,  {Borkowski} K.~J.,  {Burrows} D.~N.,   {Park} S.,  2006, \mn@doi [\apj] {10.1086/504285}, \href {https://ui.adsabs.harvard.edu/abs/2006ApJ...645..293Z} {645, 293}

\bibitem[\protect\citeauthoryear{{Zhekov}, {McCray}, {Dewey}, {Canizares}, {Borkowski}, {Burrows}  \& {Park}}{{Zhekov} et~al.}{2009}]{2009ApJ...692.1190Z}
{Zhekov} S.~A.,  {McCray} R.,  {Dewey} D.,  {Canizares} C.~R.,  {Borkowski} K.~J.,  {Burrows} D.~N.,   {Park} S.,  2009, \mn@doi [\apj] {10.1088/0004-637X/692/2/1190}, \href {https://ui.adsabs.harvard.edu/abs/2009ApJ...692.1190Z} {692, 1190}

\bibitem[\protect\citeauthoryear{{Zoglauer} et~al.,}{{Zoglauer} et~al.}{2015}]{2015ApJ...798...98Z}
{Zoglauer} A.,  et~al., 2015, \mn@doi [\apj] {10.1088/0004-637X/798/2/98}, \href {https://ui.adsabs.harvard.edu/abs/2015ApJ...798...98Z} {798, 98}

\bibitem[\protect\citeauthoryear{{de Vaucouleurs} \& {Corwin}}{{de Vaucouleurs} \& {Corwin}}{1985}]{1985ApJ...295..287D}
{de Vaucouleurs} G.,  {Corwin} H.~G. J.,  1985, \mn@doi [\apj] {10.1086/163374}, \href {https://ui.adsabs.harvard.edu/abs/1985ApJ...295..287D} {295, 287}

\bibitem[\protect\citeauthoryear{{van den Bergh} \& {Kamper}}{{van den Bergh} \& {Kamper}}{1977}]{1977ApJ...218..617V}
{van den Bergh} S.,  {Kamper} K.~W.,  1977, \mn@doi [\apj] {10.1086/155719}, \href {https://ui.adsabs.harvard.edu/abs/1977ApJ...218..617V} {218, 617}

\makeatother
\end{thebibliography}




\appendix

\section{Details of Data Reduction}
\label{sec:datadetatil}

\subsection{Source Regions}
\label{subsec:extraction}

Among the targets, SN 1885 is not apparent in the X-ray regime, hence we used a source region with a diameter equal to the Half Power Diameter of NuSTAR \citep[$\sim 60\arcsec$,][]{2013ApJ...770..103H}, centred at the location (RA = $00^{\rm h}42^{\rm m}43\fs03$, dec = $+41\degr16\arcmin04\farcs5$) that is determined by earlier optical images \citep{2019ApJ...872..191S}. The nearby bright X-ray source, Swift J0042.6+4112 \citep{2017ApJ...838...47Y}, was selected as the reference object to correct the astrometric errors (mainly translational) and the systematic offsets between the FPMA and FPMB modules \citep{2010SPIE.7738E..0ZH, 2013ApJ...770..103H}. The relative shift between its centroid position in each 3 -- 78 keV NuSTAR FPM image and its reported location in the Chandra Source Catalog 2.0 \citep{2020AAS...23515405E} was used to adjust the coordinates above. Because of the fine astrometric accuracy of the Chandra telescope and previous optical observations \citep{2019ApJ...872..191S}, we suppose that our crude corrections are already sufficient for our large extraction region.

For SN 1006, only $\sim 1/3$ of the total SNR area was covered by the observations. If the distribution of $^{44}$Ti ejecta is non-uniform, as already seen in Cas A \citep{2017ApJ...834...19G}, it would introduce significant errors.

Due to the mast movements and the attitude variations, some source regions close to the detector edge might get partly out of the field of view (FoV) during the long exposure and produce incorrect time-dependent corrections. For those regions, we adjusted the size of the sub-regions divided to account for the spatial variations of the response for extended sources.

\subsection{Background Treatments}
\label{subsec:appenbkg}

In SN 1006, the background regions free from source emission are limited due to the large size of the SNR. Therefore, we also included regions within the SNR in the {\tt nuskybgd} background analysis since the interior is dominated by $\lesssim 2$ keV thermal emission, which is negligible to NuSTAR and our analysis. Similar treatments were also adopted by \citet{2018ApJ...864...85L}. To prevent the possible $^{44}$Ti decay lines from affecting the fits of nearby instrumental lines and thus the later flux measurement of themselves, we still only used the regions outside the SNR to estimate the 65, 67, 75 and 86 keV lines of the internal component. Hence, the flux of these four instrumental lines within the source region was mostly determined by scaling the best-fit normalization in one detector with typical ratios derived from previous studies, rather than through joint-fitting spectra across the FoV.

As for G1.9+0.3, the FoV including the source region is heavily affected by SL from multiple nearby sources, which also introduces more complex background components besides the standard {\tt nuskybgd} analysis. Therefore, we simply extracted spectra from nearby source-free regions in the same detector for background subtraction, as in \citet{2015ApJ...798...98Z}. Using the generated detector maps, we checked that the defined source and background regions exclude any area that spent $> 7\%$ of exposure time in another detector due to mast motions. We note that although it did not account for the spatial variation of the background and SL perfectly, this treatment will only cause small errors because the dominant underlying emission around the $^{44}$Ti decay lines is of instrumental origin and should be fairly uniform across the same detector.

\begin{figure}
\centering
\includegraphics[width=0.7\columnwidth]{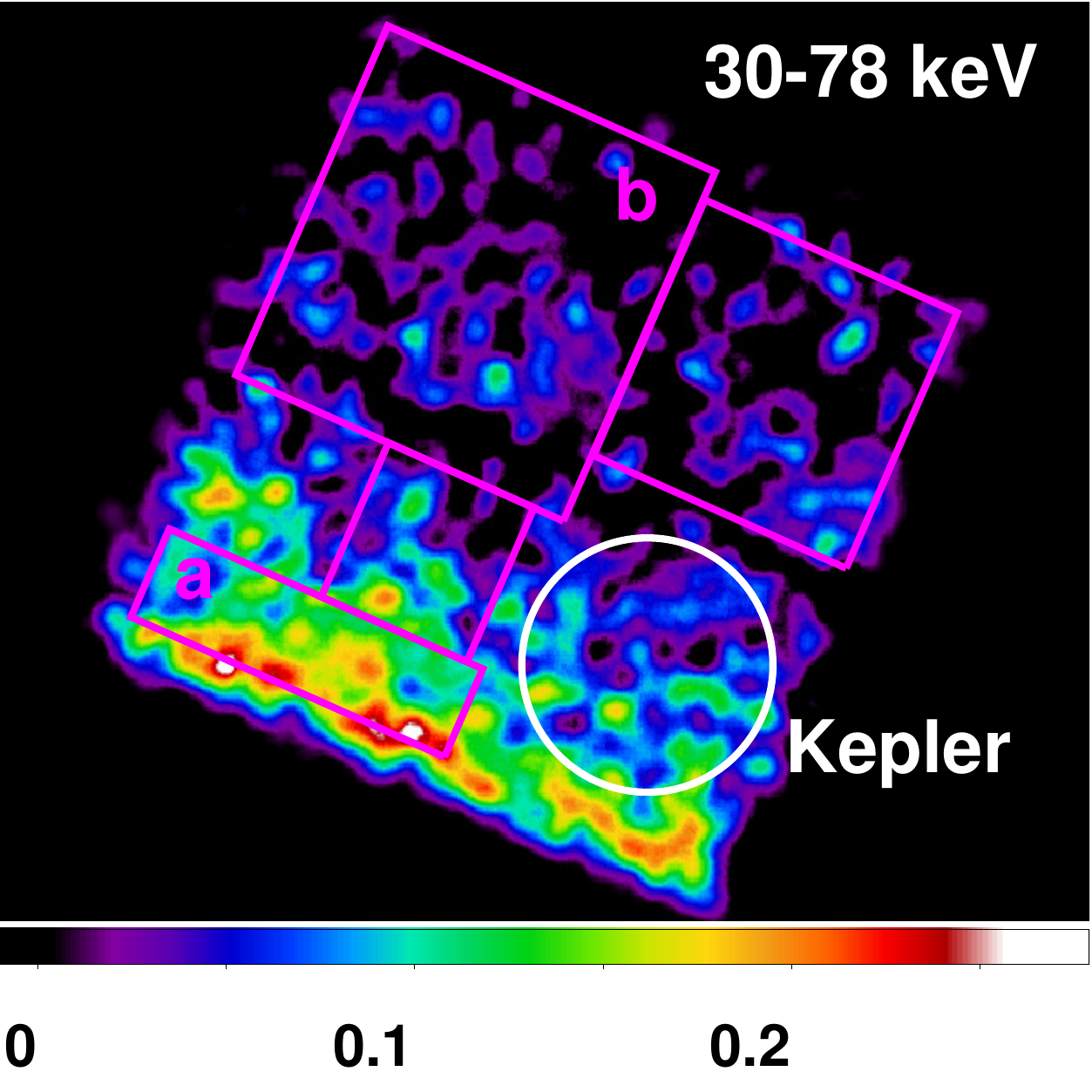}
\includegraphics[height=\columnwidth, angle=270]{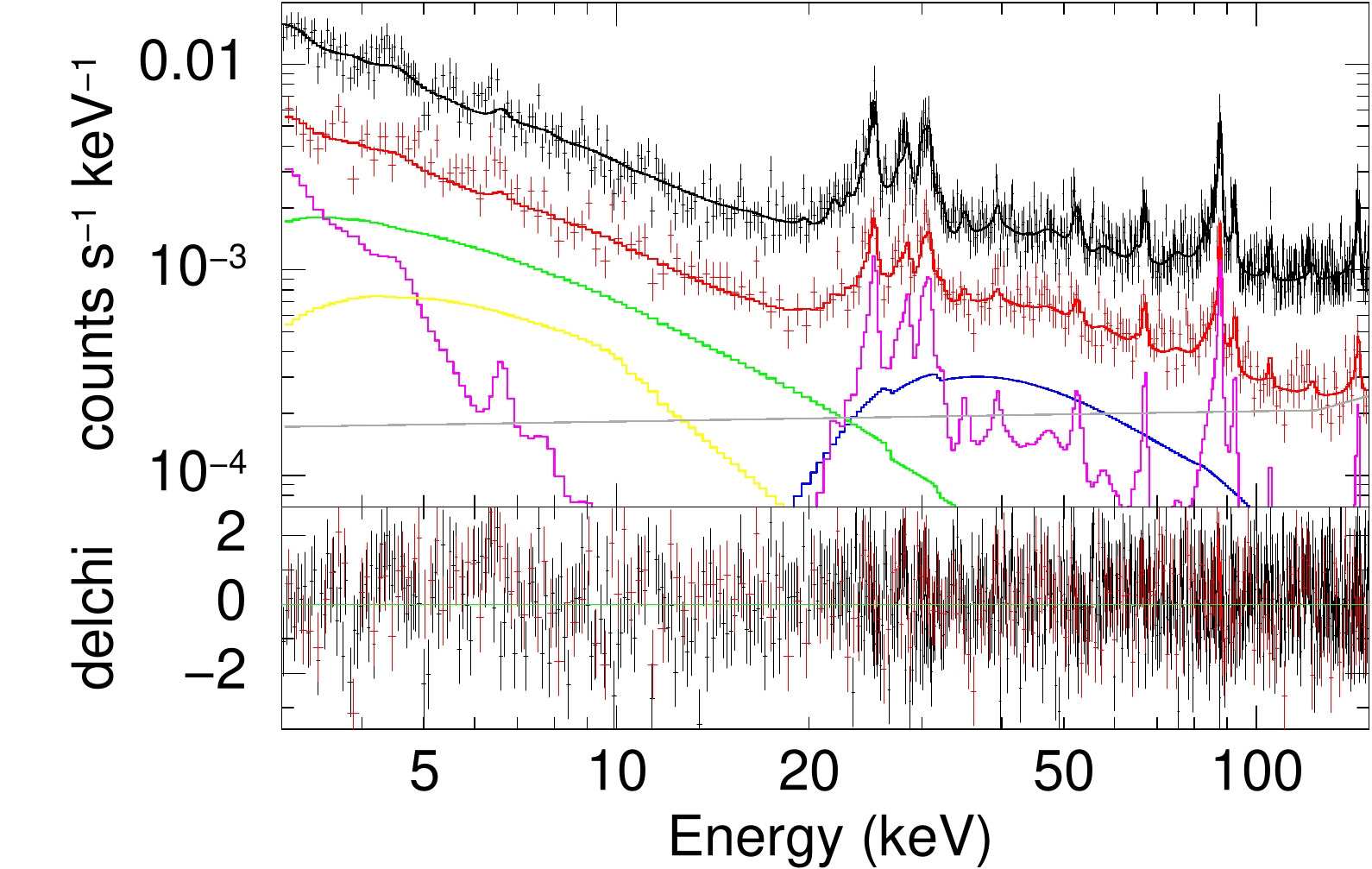}
\caption{Top: background-subtracted 30 -- 78 keV NuSTAR/FPMB count image of obs. 10801407002. The Kepler SNR and the background extraction regions are marked by white circles and the magenta rectangles, respectively. It can be seen that the source region of Kepler was affected by residual extended emission; Bottom: the spectra of background ``a" (red) and ``b" (black) and fitting residuals. The background ``a" consists of internal continuum (grey), instrumental lines plus solar emission (magenta), aperture stray light (green) and focused cosmic X-ray background (yellow). Besides, the $\sim 35$ -- $60$ keV background in region ``a" is dominated by the absorbed stray light (blue).}
\label{fig:straylight}
\end{figure}

In all NuSTAR observations of Kepler, we found residual extended emission $\gtrsim 35$ keV located at the southern half of the FoV after background subtraction by {\tt nuskybgd} (see Figure~\ref{fig:straylight} for an example). This emission is uncorrelated with the telescope position angle and is not concentrated around some well-defined sky coordinates, which indicates that it is unlikely instrumental or from the focused celestial objects. We suggest that this emission excess originates from ASL \citep{2017JATIS...3d4003M}. Although the verification of its origin requires ray-trace simulations, we note that its low energy absorption feature is consistent with the ASL spectrum of Crab \citep{2017JATIS...3d4003M}. Further discussions and evidence from an outburst event during the observations can be found in Appendix~\ref{sec:straylight}.

The accurate modelling of the ASL passing through the aperture module is intricate and beyond the scope of this paper. Therefore, we simply used a highly absorbed power law model ({\it phabs*powerlaw}) to describe the ASL spectra present in the background regions, with the photon indices tied together but the column densities varying freely to account for different depths of transmission through the aperture. The response file assigned to this model component was constructed by {\tt nuskybgd}, which excludes the optics response compared with the standard files for the focused observations. Similar treatments are also adopted for the SL analysis in \citet{2017ApJ...841...56M} and \citet{2021ApJ...909...30G}. An example of the ASL fits in the background modelling is displayed in Figure \ref{fig:straylight}. The photon indices of ASL range from 1.58 to 2.35 in the observations with ASL affecting Kepler. There can be several reasons for the difference in photon indices. These include different background regions being illuminated by different ASL sources, being blocked by varying depths of aperture absorption, state transitions in ASL sources, as well as limitations due to low counts in the spectra. It is noted that the ASL was modelled by an SL response plus a photoelectric absorption model ({\it phabs}) originally designed for the interstellar medium, which might also hinder us from obtaining an accurate photon index. However, the values of the $\chi^{2}$/d.o.f. ($\sim 1.1$) and the residual plots showed that this model is sufficient to describe the ASL spectral shape from a statistical perspective. Our analysis here focuses on the measurements of the 68 and 78 keV line fluxes rather than obtaining precise physical interpretations of the underlying ASL.

\begin{figure}
\centering
\includegraphics[width=0.8\columnwidth]{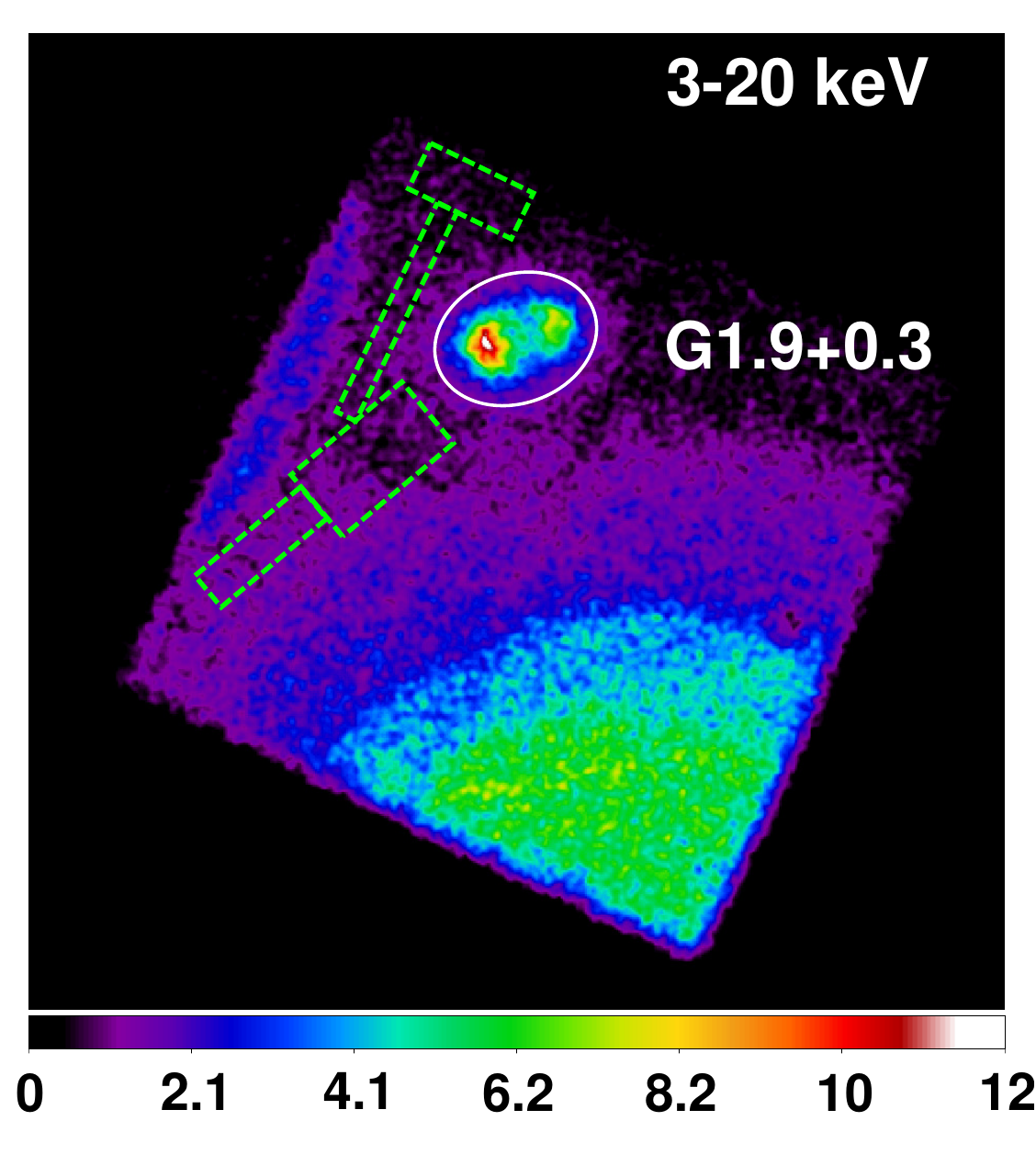}
\caption{3 -- 20 keV NuSTAR/FPMA count image of obs. 40702003008. The SNR G1.9+0.3 and the extraction
regions to fit ASL are marked by white ellipse and the green dashed rectangles, respectively.}
\label{fig:g1d9straylight}
\end{figure}

To investigate the system errors arising from the ASL spectral shape discussed in Section~\ref{subsec:error}, we conducted background analyses of the longest observations of G1.9+0.3 (obs. 40001015007 and 40702003008) in 2013 and 2021. The background regions were selected to avoid the SL illuminated regions as indicated by the StrayCats catalogue (see Figure~\ref{fig:g1d9straylight} for an example). The resulting photon indices were 2.30 and 2.73, respectively. However, complex SL patterns exist which cannot be unambiguously identified by the catalogue, and both the SL and ASL appeared to cover the entire FoV. Consequently, we obtained a high fitting $\chi^{2}$/d.o.f. value (1.36 and 1.46, respectively), and were unable to identify completely SL or ASL free background regions that could be used as a reference to provide better constraints. Photon indices within 2 -- 3 yielded similarly acceptable goodness of fit.

\section{ASL from an Nearby Transient}
\label{sec:straylight}

Confirming ASL requires ray-trace simulations to predict its pattern on the detector, and is more challenging due to the scarcity of high-energy photons compared to identifying SL sources. Nevertheless, during a nearby outburst event, a distinctive ASL pattern can emerge that is only present in the corresponding period, which can be easily linked to the transient source.

\begin{figure*}
\centering
\includegraphics[width=\textwidth]{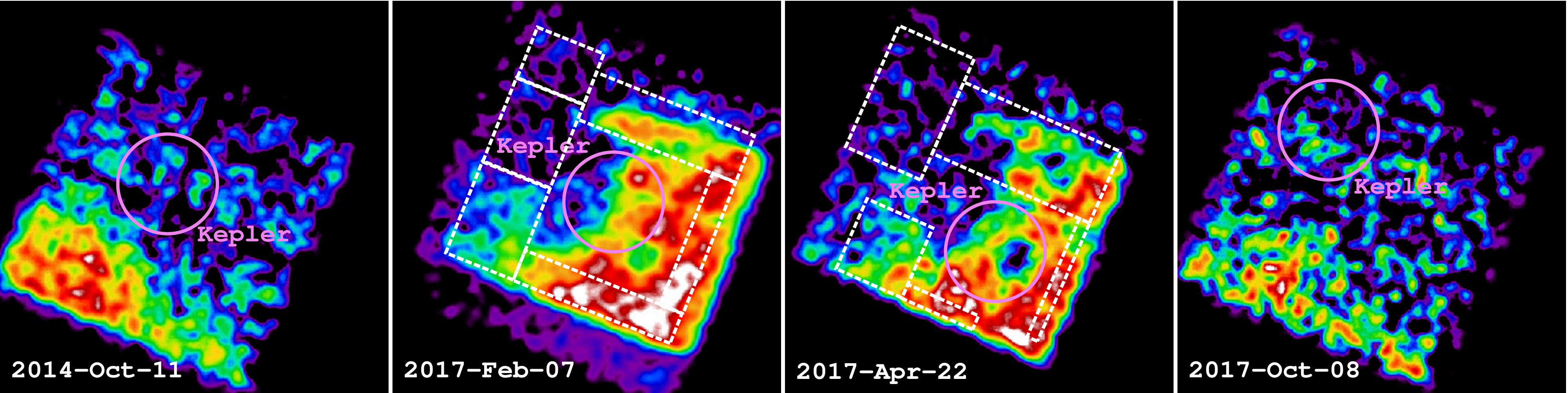}
\caption{30 -- 78 keV FPMA image of obs. 40001020002, 90201021002, 90201021004 and 90201021006 (from left to right) after standard background subtraction. The violet circles depict the source region of Kepler and the dashed white rectangles mark the extraction regions for ASL analysis. The observation start dates are at the lower left corner. These images use different scales for colour contrast.}
\label{fig:straylightevolution}
\end{figure*}

Figure~\ref{fig:straylightevolution} displays the images after background subtraction using the {\tt nuskybgd} routine for four consecutive observations of Kepler. The first and last images exhibit a pattern of residual extended emission above 30 keV similar to that shown in Figure~\ref{fig:straylight}. However, during early 2017, its shape changed significantly, accompanied by a notable increase in its intensity. This peculiar behaviour suggests an outburst event occurring in the southwest direction of Kepler.

\begin{figure}
\centering
\includegraphics[width=0.9\columnwidth]{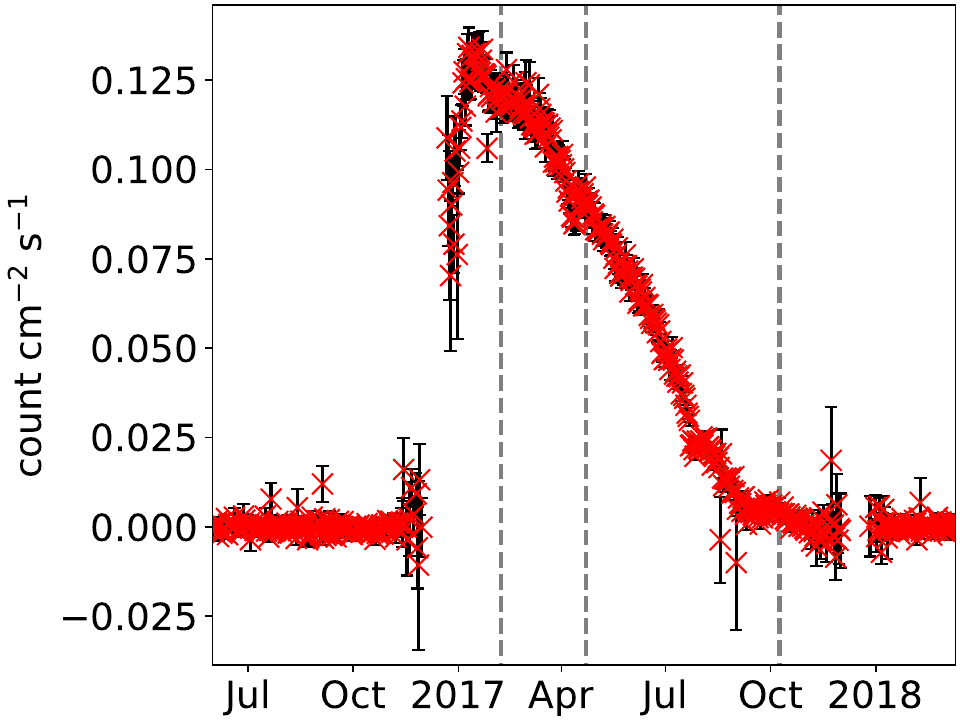}
\caption{{\it Swift}/BAT light curve of Granat 1716-249. The vertical grey dashed lines correspond to the start times of the observation of obs. 90201021002, 90201021004 and 90201021006 (from left to right).}
\label{fig:grs}
\end{figure}

We identified a black hole binary named Granat 1716-249 in the {\it Swift}/BAT Transient Monitor catalogue \citep{2013ApJS..209...14K} which underwent an outburst during this period (see Figure~\ref{fig:grs}). It returned to the quiescent state in October 2017 and the corresponding extended residual also vanished during the NuSTAR observation at the same time (see Figure~\ref{fig:straylightevolution}). Granat 1716-249 \citep[RA = $17^{\rm h}19^{\rm m}36\fs92$, dec = $-25\degr01\arcmin04\farcs12$,][]{2021A&A...649A...1G}, positioned $\sim 4.3$ degrees southwest of Kepler, falls within the typical angular range for ASL sources \citep{2017JATIS...3d4003M}. All the evidence strongly suggests that the residual emissions observed after background subtraction in obs. 90201021002 and 90201021004 are attributed to the ASL originating from Granat 1716-249.

To test if our treatment of ASL in Section~\ref{sec:data} and \ref{sec:results} is correct, we also performed the same spectral fits ({\it phabs*powerlaw} model plus the SL response) in the background regions depicted in Figure~\ref{fig:straylightevolution}. We obtained excellent goodness of fit, and the 90\% confidence interval of the photon index is 2.11 -- 2.12 and 1.87 -- 1.94 in obs. 90201021002 and 90201021004, respectively. The index difference is likely due to the different extraction regions and thus different absorptions which were poorly constrained in our model, but does not indicate the state transition. The photon index is softer than the observed index of 1.5 -- 1.8 in the spectral evolution captured by {\it Swift} during the outburst \citep{2019MNRAS.482.1587B}. This implies that our crude treatment still cannot fully reproduce accurate parameters of the ASL, although statistically sufficient in spectral fits.

We also obtained varying absorption depth across the ASL, but the errors are too large and we cannot find the crescent pattern originating from the overlapping aperture \citep{2017JATIS...3d4003M}.

\section{Source Spectra}
\label{sec:spectra}

\begin{figure*}
\centering
\includegraphics[angle=270, width=0.45\textwidth]{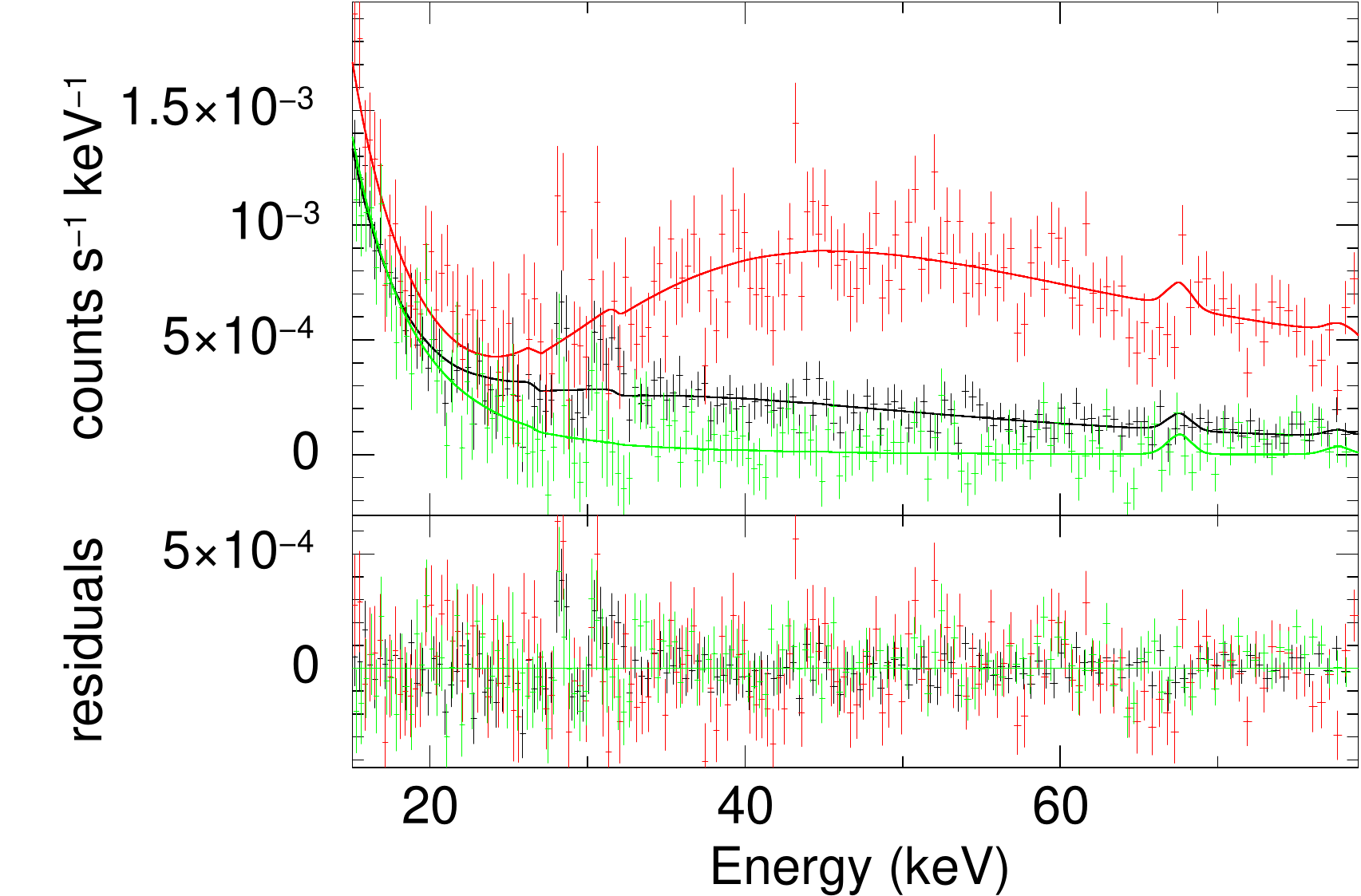}\quad\includegraphics[angle=270, width=0.45\textwidth]{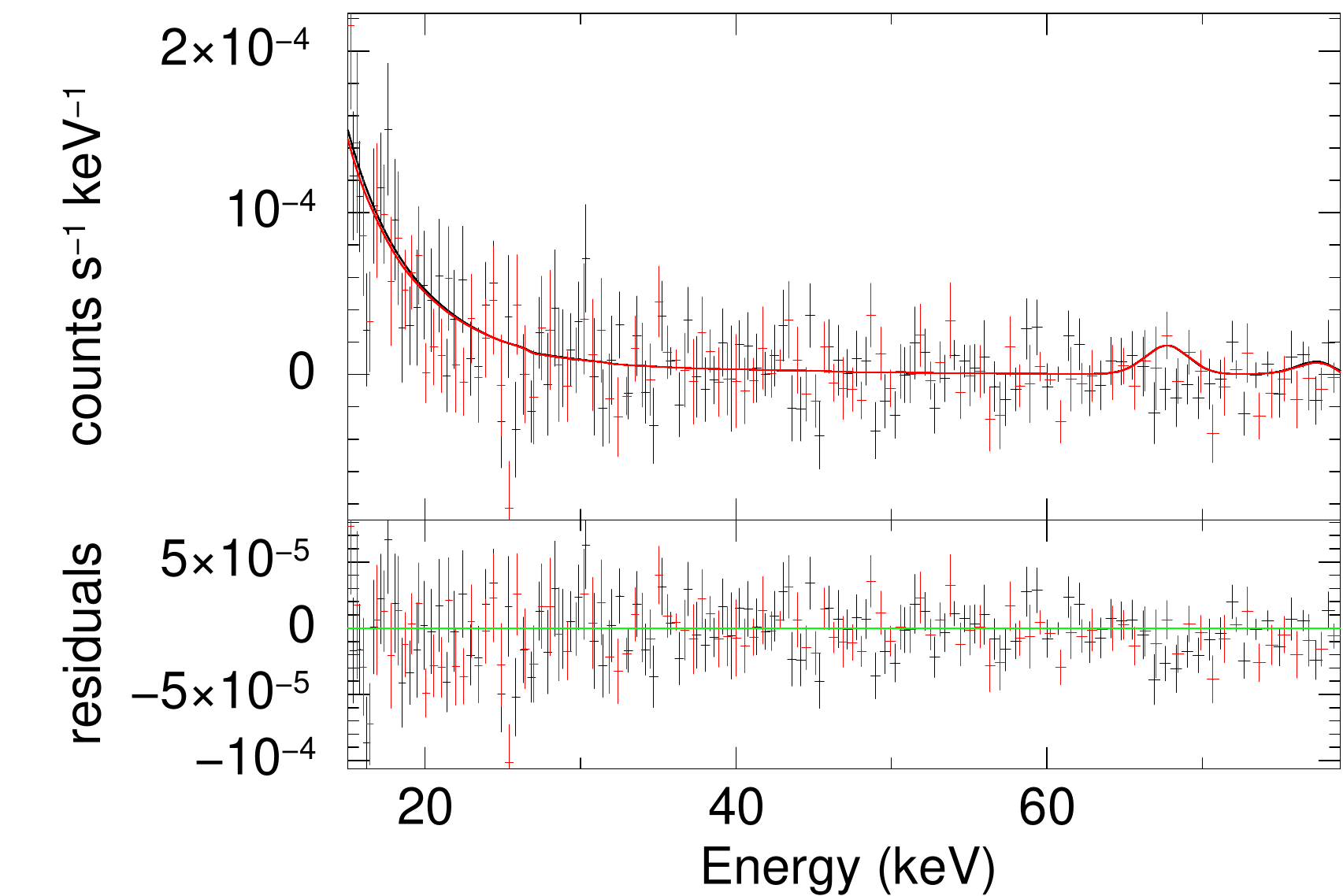}
\\
\includegraphics[angle=270, width=0.45\textwidth]{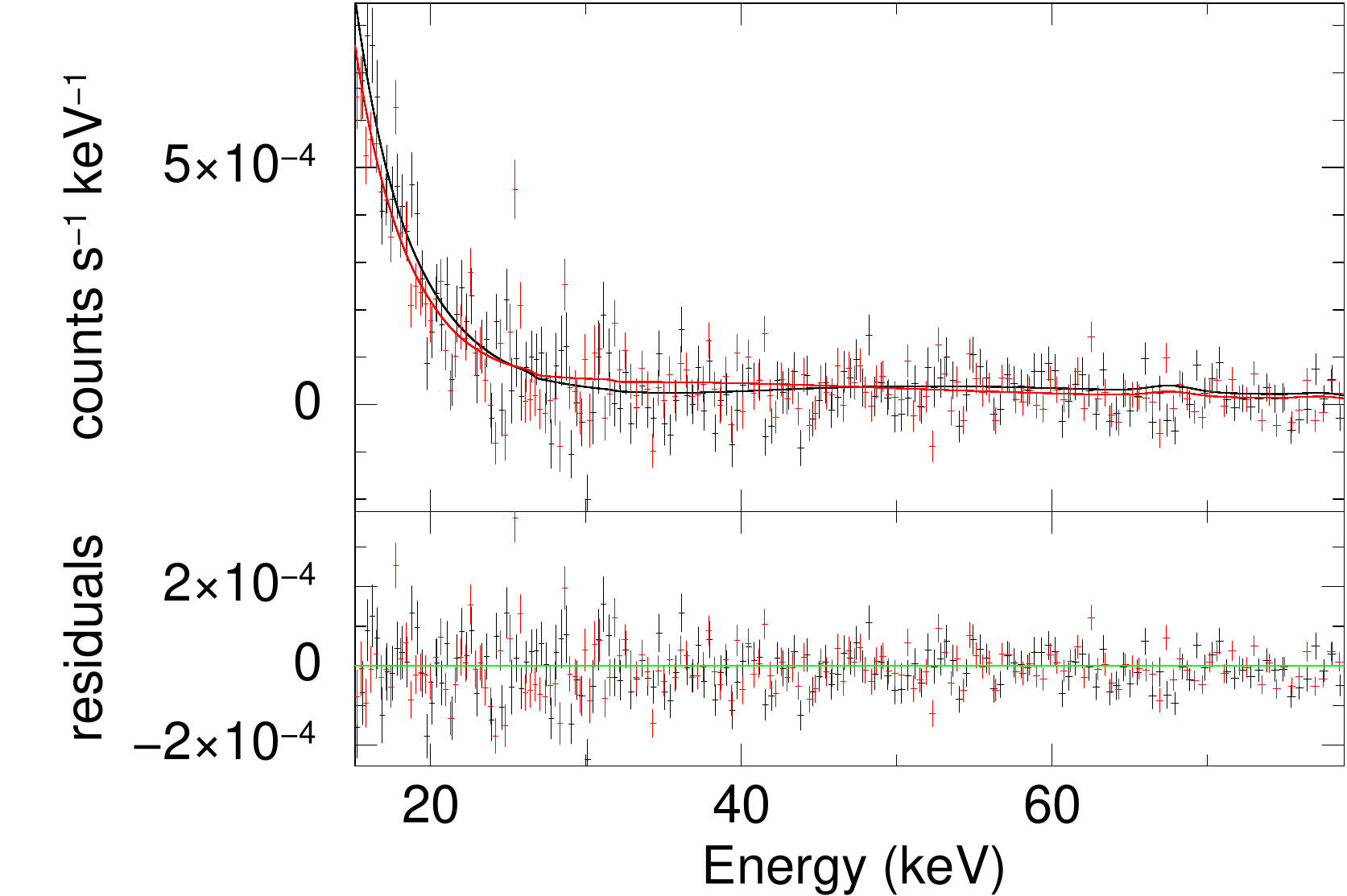}\quad\includegraphics[angle=270, width=0.45\textwidth]{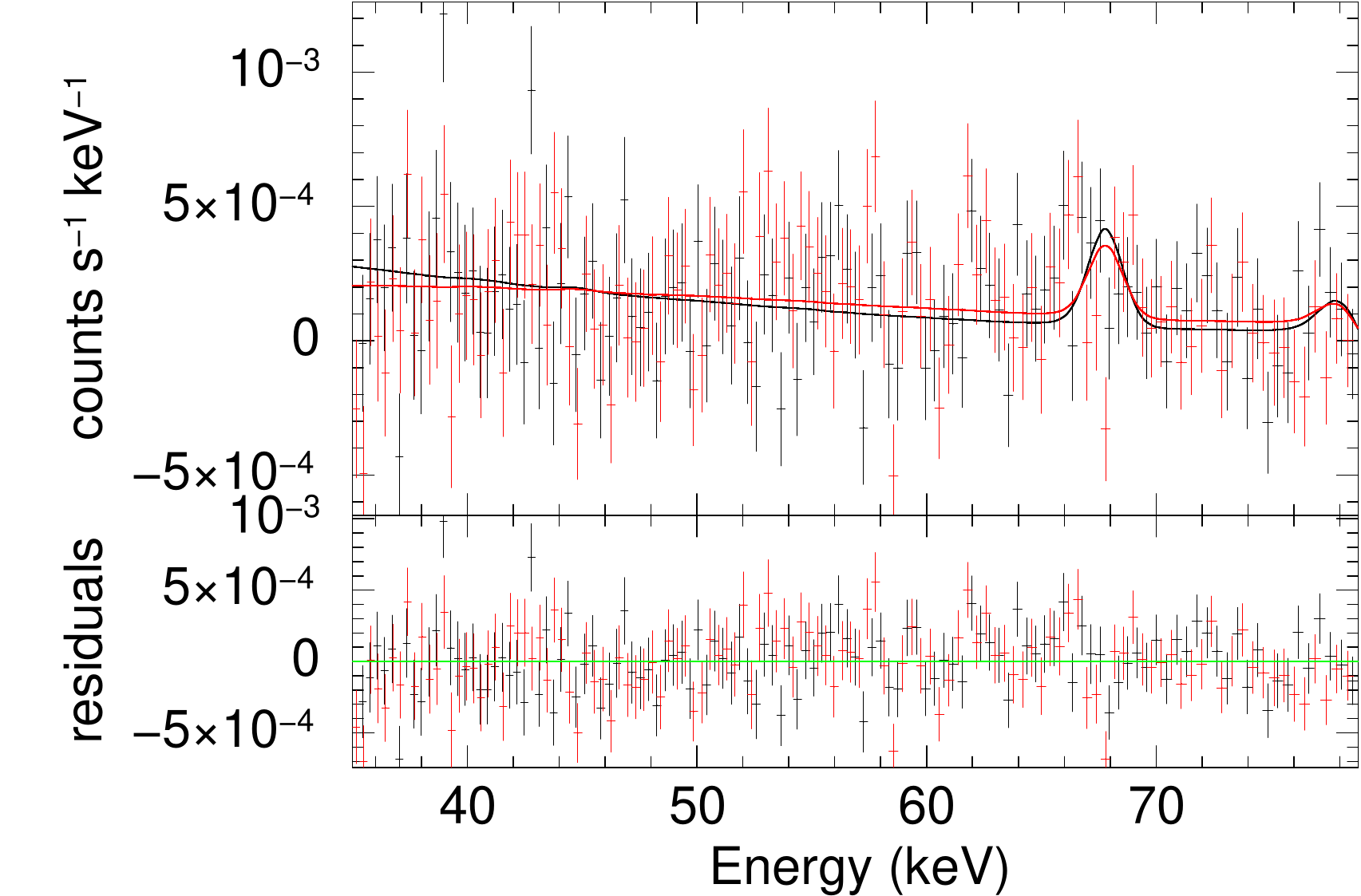}
\caption{Illustrative spectra of Kepler, SN 1885, G1.9+0.3 and SN 1006 (from top to bottom, left to right) with the $^{44}$Ti 68 and 78 keV line flux set at the 2$\sigma$ upper limits from Section \ref{sec:results}. The spectra from different observations and telescopes were co-added according to the following criteria. Kepler: whether the source region was affected by ASL from Granat 1716-249 outburst (red), constant ASL (black) or not (green); SN 1885: FPMA (black) and FPMB (red); G1.9+0.3: years 2013 (black) and 2021 (red); SN 1006: NE (black) and SW (red) segments. For the sole purpose of plotting, the spectra have been rebinned according to the optimal binning template derived from the deepest FPMA spectrum (the deepest FPMB spectrum included for SN 1885).}
\label{fig:spectra}
\end{figure*}

Figure~\ref{fig:spectra} shows the fitting of the SNR spectra with one of the background realisations (except G1.9+0.3) as an example.

\bsp	
\label{lastpage}
\end{document}